\documentclass[11pt,a4paper]{article}
\usepackage[T1]{fontenc}
\usepackage[utf8]{inputenc}
\usepackage{lmodern}
\usepackage[margin=25mm]{geometry}
\usepackage{amsmath,amssymb,mathtools,bm}
\usepackage{slashed}
\usepackage{mathrsfs}
\usepackage{booktabs,array}
\usepackage{microtype}
\usepackage{xcolor}
\usepackage{tikz}
\usetikzlibrary{arrows.meta}
\usepackage{cite}
\usepackage{hyperref}
\hypersetup{colorlinks=true,linkcolor=blue!45!black,citecolor=blue!45!black,urlcolor=blue!45!black,pdftitle={Finite-N holography and ensembles at the Leigh-Strassler fixed point},pdfauthor={Jesse van Muiden}}
\numberwithin{equation}{section}
\newcommand{\rme}{\mathrm e}
\newcommand{\dd}{\mathrm d}
\newcommand{\rmd}{\mathrm d}
\newcommand{\rmi}{\mathrm{i}}
\DeclareMathOperator{\PE}{PE}

\title{\fontsize{20pt}{24pt}\selectfont\textbf{Holographic ensembles and giant gravitons\\at the Leigh--Strassler fixed point}\vspace{2mm}}

\author{
\large{\href{mailto:jvanmuid@ic.ac.uk}{Jesse van Muiden}$^1$ and \href{mailto:}{De-liang Zhong}${}^{2,3}$}\\[5mm]
{}$^1${Abdus Salam Centre for Theoretical Physics, Imperial College London}\\
{\normalsize Prince Consort Road, London SW7 2AZ, UK}\\[5mm]
{}$^2${Institute of Theoretical Physics, Chinese Academy of Sciences}\\
{\normalsize Beijing 100190, China}\\[5mm]
{}$^3${School of Physical Sciences, University of Chinese Academy of Sciences}\\
{\normalsize Beijing 100049, China}\\[5mm]
}

\date{}

\begin{document}
\maketitle
\begin{abstract}
Finite-rank observables in holographic quantum field theories provide stringent tests of string and M-theory, but remain challenging to compute when supersymmetry is non-maximal. In this paper, we study the finite rank partition function of the $\mathcal N=1$ Leigh-Strassler CFT, which admits an interacting Fermi-gas description. We compute the partition function explicitly in two ensembles: the canonical ensemble at fixed rank $N$, and the grand canonical ensemble, where $N$ is summed over while its conjugate chemical potential $\mu$ is held fixed. In line with recent proposals, we find that the perturbative large-$N$ and large-$\mu$ expansions in both ensembles are reproduced by the on-shell action and one-loop determinant over Kaluza--Klein modes of type IIB supergravity, with appropriate ensemble-dependent boundary terms. Furthermore, the one-loop partition function of a D3-brane giant graviton precisely matches the leading nonperturbative correction to the boundary partition functions.
\end{abstract}
\clearpage
\tableofcontents
\clearpage

\section{Introduction}
\label{sec:introduction}

Finite-$N$ observables probe aspects of holography beyond perturbative supergravity. While the perturbative $1/N$ expansion captures corrections around a given gravitational saddle, non-perturbative contributions of order $\rme^{-N}$ require additional saddles, including string or brane instantons whose worldvolume theories can themselves be quantized semiclassically. Recent successful holographic matches involving quantized string and brane instantons~\cite{Gautason:2023igo,Beccaria:2023ujc,Kurlyand:2026yke,Lin:2026uzp}, as well as giant gravitons~\cite{Arai:2019xmp,Arai:2020uwd,Imamura:2021ytr,Gautason:2023igo,Lee:2023iil,Eleftheriou:2023jxr,Beccaria:2023sph,Beccaria:2023cuo,Gautason:2024nru,Beccaria:2024vfx,Lee:2024hef,Eleftheriou:2025lac,Deddo:2025lfm}, have often focused on highly symmetric backgrounds, such as AdS$_4\times S^7/\mathbf{Z}_k$, AdS$_5\times S^5$, and AdS$_7\times S^4$. Similar bulk calculations extend to less symmetric backgrounds~\cite{Gautason:2025per,Kurlyand:2026gxi,Deddo:2026exw,BenettiGenolini:2026gjn}, but exact boundary results against which to test them are often unavailable.

Our first motivation is to provide such a test in a less supersymmetric setting. We study the holographic duality between the $\mathcal N=1$ Leigh-Strassler (LS) CFT \cite{Leigh:1995ep}, and type IIB string theory on the Pilch-Warner background \cite{Pilch:2000ej,Pilch:2000fu}. Despite its reduced supersymmetry, the LS theory admits a supersymmetric $S^3\times S^1$ partition function that can be studied at finite $N$. The corresponding superconformal index has a matrix-integral representation, from which the partition function is obtained by including the supersymmetric Casimir factor~\cite{Assel:2014paa}. Remarkably, this matrix integral admits a Fermi-gas description. In a suitable limit of parameter space, this reduces to the free Fermi-gas studied in the context of the Schur limit of the $S^3 \times S^1$ partition function of $\mathcal N=4$ SYM \cite{Bourdier:2015wda}. Subsequently it can be expanded in a particularly simple giant graviton expansion, similar to results already known for $\mathcal N=4$ SYM, see for example \cite{Arai:2020qaj,Gaiotto:2021xce,Murthy:2022ien,Beccaria:2024szi} and references within.

Our second motivation is to test the recent proposal that top-down holographic dualities are naturally formulated in terms of grand canonical ensembles of boundary theories~\cite{Gautason:2025plx}. The Fermi-gas description provides a convenient framework for this investigation.\footnote{The Fermi-gas description has shown to be particularly useful in the context of the ABJM theory in \cite{Marino:2011eh}, and gives a direct method to test the grand canonical formulation of bulk string and M-theory, but it is not essential for the bulk analysis.} In this description $N$ plays the role of particle number. Passing to the grand canonical ensemble allows $N$ to fluctuate while holding its conjugate chemical potential $\mu$ fixed, considerably simplifying the analysis in the free-fermion limit. Although summing over boundary theories with different $N$ may seem unconventional, it has a natural bulk interpretation: instead of fixing the flux number and the associated number of branes, one fixes the boundary holonomy of the associated gauge potential, which is identified with the boundary chemical potential $\mu$.\footnote{To implement the choice of ensembles in eleven dimensions one is to start from a parent theory, with two fields
\begin{equation*}
    S_{11d}[A_3,H_7] = \frac{1}{2\kappa^2} \int - R \star 1 + \frac12 H_7 \wedge \star H_7 + \rmi \, \rmd A_3 \wedge H_7 + \frac{\rmi}{6} A_3 \wedge \rmd A_3 \wedge \rmd A_3\,,
\end{equation*}
which upon integrating out $H_7$ gives the standard eleven-dimensional action in terms of $A_3$, but when integrating out $A_3$ it provides a boundary term and can induce additional non-local interactions.} Importantly, the drastic simplification in the free fermi description in the grand canonical ensemble of the boundary theory implied in the corresponding supersymmetric M-theory bulk that the supergravity and membrane instanton saddles are 1-loop exact \cite{Gautason:2025plx}.

The ensemble proposal was inspired by a worldsheet approach in AdS$_3$ \cite{Kutasov:1999xu,Kim:2015gak,Eberhardt:2020bgq,Aharony:2024fid}. It was later argued, more generally, to be the natural approach for semi-classical M2-branes and for effective supergravity \cite{Gautason:2025per,Gautason:2025plx}. For the membrane it has been tested to two-loop order \cite{Beccaria:2026zjj,Tseytlin:2026dhy}. In the simplest setups, as in ours, the measure of the ensemble changing integral is trivial; in general it does depend on the topology of the holographic boundary \cite{Bobev:2026gir}. At the supergravity level, the grand canonical proposal has been tested to the eighth derivative order in M-theory \cite{Bobev:2026gir,BenettiGenolini:2026cyc} and at one-loop order \cite{Bobev:2026gir,BenettiGenolini:2026kjy,Kurlyand:2026cyz}. In type IIB string theory it has been tested  at the level of the two-derivative supergravity action, its 1-loop correction and D-brane instantons \cite{vanMuiden:2026lno}. 

For the IIB setup there is a subtlety arising from the self-duality constraint on the five-form flux, which we will encounter here as well. It can be dealt with consistently by adding topological terms to the Pasti--Sorokin--Tonin (PST) action of IIB supergravity \cite{Kurlyand:2022vzv,Adhikari:2026rfb}, with appropriate couplings that fix the boundary ensemble \cite{vanMuiden:2026lno,Adhikari:2026xdb}. As in eleven dimensions, one can start from a parent action in which the four-form gauge potential $C_4$ and an auxiliary five-form are independent. The difference is that in the PST formulation of the IIB effective action, as summarised in section \ref{sec:supergravity}, the topological term $\alpha S_{\rm top}$ contributes non-trivially on the boundary in both ensembles, with $\alpha=-1$ at fixed electric holonomy and $\alpha=+1$ at fixed D3 Page charge. Determining these boundary terms directly from type IIB string theory, within for example the effective-action approach of \cite{Fradkin:1984pq,Fradkin:1985fq,Tseytlin:1988tv}, remains an open problem, despite recent progress on deriving gravitational boundary terms from the string worldsheet \cite{Ahmadain:2024uom,Ahmadain:2024uyo}. 

To circumvent this subtlety, we instead adopt entirely a low-energy supergravity approach and compute the bulk partition function at fixed boundary holonomy of the gauge field
\begin{equation}
	Z_{\text{SUGRA}}[\left. C_p\right|_\partial] = \int \mathcal D[g,C_p,\ldots] \, \rme^{-S_{\text{SUGRA}}(g,\ldots)} \,,
\end{equation}
where the bulk action is a functional of $p$-form gauge potentials $C_p$ of degrees $0 \leq p \leq 4$. We fix their holonomy on the boundary of AdS, not the dual flux $\star \rmd C$ on the internal space. The ensemble at fixed flux follows by integrating over the boundary potential\footnote{Chern-Simons terms can alter the canonically conjugate symplectic pairs of gauge potential and dual flux.}
\begin{equation}
	Z_{\text{SUGRA}}[\left. \star \rmd C_p\right|_\partial] = \int \mathcal D[\left.C_p \right|_\partial] \,Z_{\text{SUGRA}}[\left. C_p\right|_\partial] \, \rme^{\int_\partial C_p \wedge \star \rmd C_p}\,.
\end{equation}
We apply this method to the four-form gauge potential in the Pilch-Warner background \cite{Pilch:2000ej}, a deformation of AdS$_5\times S^5$ dual to the LS CFT~\cite{Leigh:1995ep}. We compute the supergravity on-shell action in both ensembles, at fixed flux number $N$ and fixed conjugate boundary holonomy $\mu$, and determine the one-loop contributions from Kaluza--Klein (KK) fluctuations, using the results of \cite{Bobev:2020lsk} for the massive KK modes. Finally, we compute the leading exponentially suppressed contributions, for an unrefeined limit of the index, in the large-$N$ and large-$\mu$ expansions, coming from D3-brane giant gravitons.

The Fermi-gas description of the supersymmetric LS partition enables a nonperturbative comparison between the bulk and boundary theories in the large-$N$ and large-$\mu$ expansions. In the unrefined limit, a redefinition of fugacities maps the LS partition function to the supersymmetric Schur partition function of the $\mathcal N = 4$ theory. The grand canonical potential of the unrefined LS index equals\footnote{As we will explain, the equality also holds with an additional flavor fugacity turned on.}
\begin{equation}
	 \mathcal J(\mu) = \frac{\rmi \mu^2}{\pi \tau} + \frac{\rmi\pi}{4\tau} +  \frac{\rmi}{\pi\tau} \sum_{n=1}^{\infty}\frac1n\binom{2n}{n} \bigl(H_{2n-1}-H_n-\mu\bigr)e^{-2n\mu}\,.
\end{equation}
The LS partition function is thus related to the Schur partition function of the $\mathcal N=4$ theory. Similar relations were observed before, for theories that are connected by a supersymmetric adjoint mass deformation \cite{Gadde:2010en,Kang:2021lic}. This relation is manifest in the matrix model, but from the bulk point of view we find it highly surprising. The deformed background has a different on-shell action, KK-spectrum, and the non-perturbative D3-branes wrap deformed squashed spheres embedded in $S^5$.
Nevertheless, after a simple redefinition of parameters, 
all their contributions to the bulk partition function equal that of the round AdS$_5 \times S^5$ geometry.

\section{The Leigh--Strassler CFT partition function}
\label{sec:qft}
The Leigh--Strassler (LS) theory is the interacting $\mathcal{N}=1$ infrared fixed point that arises from a mass deformation of the $\mathcal N=4$ SYM theory. The undeformed theory has bosonic symmetry $SO(2,4)\times SO(6)_R$. The deformation preserves an internal $SU(2)_F\times U(1)_R$ subgroup. Conformal invariance is broken along the flow and recovered at the LS fixed point. In $\mathcal N=1$ language, the theory contains a vector multiplet and three adjoint chiral multiplets $X,Y,Z$, with superpotential~\cite{Leigh:1995ep}
\begin{equation}
 W=\operatorname{Tr}\bigl(Z[X,Y]\bigr)+\frac m2\operatorname{Tr}Z^2\,, \qquad W_{\rm eff}=-\frac1{2m}\operatorname{Tr}[X,Y]^2\,,
 \label{qft:superpotential}
\end{equation}
where the effective superpotential comes from integrating out $Z$. The preserved R symmetry assigns charges $(1/2,1/2,1)$ to $(X,Y,Z)$, and the $SU(2)_F$ flavour symmetry rotates $(X,Y)$ with Cartan charges $(+1,-1)$ while leaves $Z$ invariant. 

Our goal is to compute holographically the supersymmetric $S^3\times S^1$ partition function $Z_N$ of the Leigh--Strassler theory. On the field-theory side, $Z_N$ is obtained by combining the superconformal index with the supersymmetric Casimir factor~\cite{Assel:2014paa,Assel:2015nca,ArabiArdehali:2015iow}. We first recall the superconformal index
\begin{equation}
 \begin{split} \mathcal I_N(x,y,u) &=\operatorname{Tr}_{\mathcal H_N} (-1)^{\mathsf F}e^{-\varepsilon H_Q} x^{2(2j_2+R)}y^{2j_1}u^F\,, \end{split}
 \label{qft:fugacities}
\end{equation}
where $\mathcal H_N$ is the gauge-invariant Hilbert space on $S^3$, $\mathsf F$ denotes fermion number, $j_1,j_2$ are the Cartans of the two $SU(2)$ rotations, $R$ is the preserved R charge, and $F$ is the Cartan charge of $SU(2)_F$. We choose a supercharge $Q$ with charges $(D,j_1,j_2,R_{\rm sc},F)=(1/2,0,-1/2,1,0)$. The supersymmetric Hamiltonian that grades the trace is its anti-commutator
\begin{equation}
 H_Q\equiv\{Q,Q^\dagger\}=D-2j_2-\frac32R_{\rm sc}\geq0\,,
 \label{qft:supercharge}
\end{equation}
and only states saturating the lower bound contribute to the index. Here $D$ is the dilatation operator, identified in radial quantization with the Hamiltonian on the unit-radius cylinder, and $R_{\rm sc}$ is the exact superconformal R charge. 

The fugacities $x,y,u$ are conjugate to $2(2j_2+R)$, $2j_1$ and $F$, respectively. The index can be computed with R\"omelsberger's prescription by assigning the correct IR charges to the microscopic fields in the UV~\cite{Romelsberger:2007ec}. For an adjoint chiral multiplet of scalar R-charge $r$ and flavour fugacity $u$, and for a vector multiplet, the single-letter contributions are
\begin{equation}
 f_\chi(x,y;r,u) = \frac{x^{2r}u-x^{4-2r}u^{-1}}{(1-x^2y)(1-x^2/y)}\,, \qquad f_V(x,y) = \frac{2x^4-x^2(y+y^{-1})}{(1-x^2y)(1-x^2/y)}\,.
 \label{qft:elementaryletters}
\end{equation}
Adding the contributions from the vector multiplet and the two massless chiral multiplets, with $r=1/2$ and flavour weights $u$ and $u^{-1}$, gives\footnote{The single-letter index of the massive chiral multiplet $Z$, with $R$-charge $1$ and singlet for the flavour symmetry, trivializes.}
\begin{equation}\label{qft:letter}   
\begin{aligned}
 f_{\rm LS}(x,y,u) & =  f_V(x,y) + f_\chi(x,y;\tfrac{1}{2},u)+ f_\chi(x,y;\tfrac{1}{2},u^{-1}) \\
 & = \frac{(x-x^3)(u+u^{-1})-x^2(y+y^{-1})+2x^4}{(1-x^2y)(1-x^2/y)}\,.
\end{aligned}
\end{equation}
The partition function follows by summing over the gauge invariant states built from the single-letter index $f_{LS}$, and by adding the Casimir contribution $E_{\rm susy}$
\begin{equation}
 Z_N =e^{-\mathcal F_N(x,y,u)} \int_{U(N)}\mathrm dU\, \exp\!\left[\sum_{k=1}^{\infty} \frac{f_{\rm LS}(x^k,y^k,u^k)}{k} \operatorname{Tr}U^k\operatorname{Tr}U^{-k}\right]\,,
 \label{qft:matrix}
\end{equation}
where $\mathcal F_N=\beta E_{\rm susy}$. We compute this partition function both in the \textit{canonical ensemble at fixed $N$} and in the \textit{grand-canonical ensemble at fixed chemical potential $\mu$}, where $N$ is summed over. To facilitate the analysis in both ensembles, we rewrite the matrix integral as the partition function of a Fermi gas. We parametrize the eigenvalues of $U$ by $z_i=e^{2\rmi\alpha_i}$, with $0\leq\alpha_i<\pi$, and define $x=e^{\rmi\pi\tau}$ and $u=e^{\rmi\pi\varphi}$. 
As we show below, in these variables the partition function admits the Fermi-gas representation
\begin{equation}
 Z_N(x,y,u)= e^{-\mathcal F_N(x,y,u)} \frac{x^{-N^2/4}e^{NA}}{\Delta_N(x,\varphi)} Z_{F,N}^{\rm int}(x,y,u)\,,
 \label{qft:fermiprefactor}
\end{equation}
with
\begin{equation}
 Z_{F,N}^{\rm int}(x,y,u)= \frac1{N!}\int_{[0,\pi)^N}\mathrm d^N\alpha\, \det_{i,j}\rho_{x,\varphi}(\alpha_i,\alpha_j)\, e^{-\sum_{i<j}V(\alpha_i-\alpha_j)}\, .
 \label{qft:fermicoeff}
\end{equation}
Here $\Delta_N$ is the normalization \eqref{qft:cauchyparity}, $\rho_{x,\varphi}$ is the kernel \eqref{qft:kernel}, and the diagonal interaction $A$ and the pair potential $V$ are given in \eqref{qft:interactionweight} and \eqref{qft:pairweight}, respectively.
Appendix \ref{app:specialfunctions} defines the theta functions in \eqref{app:thetaseries} and gives $e^A$ as a product in \eqref{app:diagonalproduct}.

To derive \eqref{qft:fermiprefactor}, we decompose the single-letter index into a contribution $f_0$ with a known free Fermi-gas description and a remainder $\delta f$ that generates the interactions:
\begin{equation}
 \begin{split} f_{\rm LS}(x,y,u)&=f_0(x,u)+\delta f(x,y,u)\,, \end{split}
 \label{qft:lettersplit}
\end{equation}
where
\begin{equation}
 f_0(x,u) = \frac{x(u+u^{-1})-2x^2}{1-x^2}\,, \quad   \delta f(x,y,u) = - \frac{y+y^{-1}-2}{(1-x^2y)(1-x^2/y)} \frac{x^2(1-xu)(1-x/u)}{1-x^2}\,.
 \label{qft:interaction}
\end{equation}
Up to a redefinition of fugacities, $f_0$ coincides with the single-letter Schur index of $\mathcal N=4$ SYM:
\begin{equation}
 f_{\rm LS}(x,1,u)=f_0(x,u) =f_{\mathcal N=4,\,\mathrm{Schur}}(x^2,u)\,,
 \label{qft:schur}
\end{equation}
whose free Fermi-gas description was given in~\cite{Bourdier:2015wda}\footnote{See also \cite{Bourdier:2015sga} for relevant generalisations.}. 

To write the free density matrix explicitly, we set $\alpha_{ij}=\alpha_i-\alpha_j$ and $T_k=\operatorname{Tr}U^k=\sum_i e^{2\rmi k\alpha_i}$. In these conventions, the free-gas result of~\cite{Bourdier:2015wda} writes the matrix integral as
\begin{equation}
 \mathcal W_{0,N}(\boldsymbol\alpha) = \frac1{\pi^N}\prod_{i<j}4\sin^2\alpha_{ij}\, \exp\!\left[\sum_{k=1}^{\infty}\frac{f_0(x^k,u^k)}{k}T_kT_{-k}\right] = \frac{x^{-N^2/4}}{\Delta_N(x,\varphi)} \det_{i,j}\rho_{x,\varphi}(\alpha_i,\alpha_j)\,,
 \label{qft:cauchy}
\end{equation}
where the one-particle density matrix on $[0,\pi)$ has kernel
\begin{equation}
 \rho_{x,\varphi}(\alpha,\alpha') = \frac{\vartheta_2(0,x)\vartheta_4(0,x)}{2\pi} \frac{\vartheta_2(\alpha-\alpha'+\tfrac{\pi\varphi}{2},x)}{\vartheta_4(\alpha-\alpha'+\tfrac{\pi\varphi}{2},x)}\,,
 \label{qft:kernel}
\end{equation}
and the rank-dependent normalization is
\begin{equation}
 \Delta_N(x,\varphi)= \begin{cases}
 \vartheta_3(\pi N\varphi/2,x)/\vartheta_3(0,x)\,,&N\text{ even}\,,\\[2pt]
 \vartheta_2(\pi N\varphi/2,x)/\vartheta_3(0,x)\,,&N\text{ odd}\,.
 \end{cases}
 \label{qft:cauchyparity}
\end{equation}

It remains to determine the contribution $\exp[\sum_{k\geq1}d_kT_kT_{-k}/k]$ from $\delta f$, where $d_k=\delta f(x^k,y^k,u^k)$. We can separate the diagonal terms as
\begin{equation}
 T_kT_{-k}=\sum_{i,j}e^{2\rmi k\alpha_{ij}} = N+2\sum_{i<j}\cos(2k\alpha_{ij})\,.
 \label{qft:tracemodes}
\end{equation}
such that
\begin{equation}
 \begin{split} \sum_{k=1}^{\infty}\frac{d_k}{k}T_kT_{-k}  = NA-\sum_{i<j}V(\alpha_{ij})\,, \quad \text{where} \quad A = \sum_{k=1}^{\infty}\frac{d_k}{k}\,, \quad V(a) = -2\sum_{k=1}^{\infty}\frac{d_k}{k}\cos(2ka)\,. \end{split}
 \label{qft:interactionweight}
\end{equation}
The potential also has a direct interpretation in terms of localized one-loop determinants. With the Casimir factor subtracted, the chiral multiplet contributions on $S^3\times S^1$ are elliptic gamma functions~\cite{Assel:2014paa}, and they generate the potential
\begin{equation}
 e^{-V(a;x,y,u)} = \frac{\mathcal P(a;x,y,u)}{\mathcal P(a;x,1,u)}\,,
 \label{qft:pairweight}
\end{equation}
where the denominator subtracts the free-fermion contribution and
\begin{equation}
 \mathcal P(a;x,y,u) = \prod_{\epsilon=\pm1} \frac{\Gamma_e\!\left(xu\,e^{2\rmi\epsilon a};x^2y,x^2/y\right)\Gamma_e\!\left(xu^{-1}e^{2\rmi\epsilon a};x^2y,x^2/y\right)}{\Gamma_e\!\left(e^{2\rmi\epsilon a};x^2y,x^2/y\right)}\,,
 \label{qft:pairdeterminant}
\end{equation}
with the elliptic-$\Gamma_e$ function defined in~\eqref{app:ellipticgamma}. Combining \eqref{qft:interactionweight} with \eqref{qft:cauchy} and the Casimir factor in \eqref{qft:matrix} gives the Fermi-gas form \eqref{qft:fermiprefactor}. A similar interacting-gas structure appears in three-dimensional $\mathcal N=2$ CS-matter theories~\cite{Marino:2012az}, where the potential interactions were generated by the one-loop determinant of 3d chiral multiplets instead. 

\subsection{The supersymmetric Casimir factor}

We now compute the supersymmetric Casimir factor $e^{-\mathcal{F}_N}$ in \eqref{qft:matrix}; the result is \eqref{qft:flavouredcasimir}. We use the fugacities $x=e^{\rmi\pi\tau}$ and $u=e^{\rmi\pi\varphi}$ defined above, and we introduce similarly $y=e^{2\pi\rmi\psi}$. The periodicity of the Euclidean circle is $\beta$, and we denote by $b_1,b_2$ the two Hopf parameters fixing the complex structure of the background manifold. In terms of these parameters, the fugacities are\footnote{In the conventional Hopf-surface parametrization, $(w_1,w_2)\sim(pw_1,qw_2)$ on $\mathbb C^2\setminus\{0\}$, the complex-structure parameters are $p=x^2y=e^{-\beta b_1}=e^{2\pi\rmi(\tau+\psi)}$ and $q=x^2/y=e^{-\beta b_2}=e^{2\pi\rmi(\tau-\psi)}$~\cite{Assel:2014paa}.}
\begin{equation}
 x=e^{\rmi\pi\tau} = e^{-\beta(b_1+b_2)/4}\,,\qquad y = e^{2\pi\rmi\psi}=e^{\beta(b_2-b_1)/2}\,,\qquad u = e^{\rmi\pi\varphi}\,.
 \label{qft:potentials}
\end{equation}
The supersymmetric Casimir energy follows from the equivariant integration of the anomaly polynomial~\cite{Bobev:2015kza}, which gives
\begin{equation}
 \begin{split} E_{\rm susy} = N^2 \frac{(b_1+b_2)[(b_1+b_2)^2+16\pi^2\varphi^2/\beta^2]} {64b_1b_2}\,, \qquad \mathcal F_N=\beta E_{\rm susy}\,. \end{split}
 \label{qft:hopfcasimir}
\end{equation}
In terms of the original matrix model parameters, it becomes
\begin{equation}
 \mathcal F_N(x,y,u)=N^2 \mathcal{C}(x,y,u)\,, \qquad \mathcal{C}(x,y,u)=-\frac{\rmi\pi\tau(\tau^2-\varphi^2)} {4(\tau^2-\psi^2)}\,.
 \label{qft:flavouredcasimir}
\end{equation}
This gives the Casimir factor at arbitrary flavour fugacity. At $u=1$ it reduces to $E_{\rm susy}=N^2(b_1+b_2)^3/(64b_1b_2)$, and on the round LS background $b_1=b_2=1$ it gives $E_{\rm susy}=N^2/8$.\footnote{With matched fugacities, the $\mathcal N=4$ Schur Casimir exponent obeys $\mathcal F_N^{\mathcal N=4}=\left.\mathcal F_N^{\rm LS}\right|_{y=1}=-\rmi\pi N^2(\tau^2-\varphi^2)/(4\tau)$. The round LS background has $\beta=-2\rmi\pi\tau$, whereas the Schur representative of~\cite{vanMuiden:2026lno} has $\beta=-\rmi\pi\tau$. At $u=1$ their energies are $N^2/8$ and $N^2/4$, respectively: the matched quantity is $\beta E_{\rm susy}$.}

In the large-$N$ phase governed by a uniform holonomy distribution, the Casimir term $\mathcal F_N$ is the only perturbative $N$-dependent contribution to the free energy. The Fermi-gas representation \eqref{qft:fermiprefactor} seems to contain additional terms from the prefactor 
\begin{equation}
 \frac{x^{-N^2/4}e^{NA}}{\Delta_N(x,\varphi)}\,.
\end{equation}
However, their rank dependence cancels against contributions from $Z_{F,N}^{\rm int}$. These factors were extracted to rewrite the matrix model integral in terms of a Fermi-gas. 

We separate the full partition function into perturbative and non-perturbative contributions,
\begin{equation}
 \begin{split} Z_N=Z_N^{\mathrm p}+Z_N^{\mathrm{np}}\,,\qquad Z_N^{\mathrm p}(x,y,u)=e^{-\mathcal{C}N^2}\mathcal I_\infty(x,y,u)\,, \end{split}
 \label{qft:refinedperturbative}
\end{equation}
where $\mathcal I_\infty$ contains the contribution of protected gauge-invariant single-trace states and all their multi-trace combinations, with the prescribed supersymmetric signs~\cite{Kinney:2005ej,Bobev:2020lsk},
\begin{equation}
 \begin{split} \mathcal I_\infty &= \lim_{N\to\infty}\left[e^{\mathcal{C}N^2}Z_N\right] = \prod_{k=1}^\infty \frac{1}{1 - f_{\text{LS}}(x^k,y^k,u^k)} = \frac{(x^2y;x^2y)_\infty(x^2/y;x^2/y)_\infty} {(x^2;x^2)_\infty(xu;xu)_\infty(x/u;x/u)_\infty}\,. \end{split}
 \label{qft:infinitefluctuations}
\end{equation}
The remainder $Z_N^{\mathrm{np}}=Z_N-Z_N^{\mathrm p}$ contains all corrections that are exponentially suppressed relative to $Z_N^{\mathrm p}$ at large $N$. We will reproduce these contributions holographically using type IIB supergravity, noting that $\mathcal I_{\infty}$ was already reproduced in \cite{Bobev:2020lsk}.

\subsection{The grand canonical ensemble}

We are also interested in the bulk dual of the grand canonical ensemble of the boundary theory. In this ensemble, $N$ fluctuates and its conjugate chemical potential $\mu$ is held fixed. We derive the leading non-perturbative correction at general fugacities, \eqref{qft:leadingrefined}, and the complete partition function at $y = u = 1$, \eqref{qft:thetaensemble}. The grand canonical partition function is defined as
\begin{equation}
 \Xi(\mu;x,y,u) = \sum_{N=0}^{\infty}e^{\mu N}Z_N(x,y,u)\,, \qquad Z_0 = 1\,.
 \label{qft:refinedrankensemble}
\end{equation}
The sum converges for any value of $\mu$, as long as $\mathcal{C}>0$. For real $\mu\to+\infty$, completing the square in the exponent of $e^{\mu N} Z_N^{\mathrm p}$ determines the saddle to be at
\begin{equation}
 \mu N-\mathcal{C}N^2=\frac{\mu^2}{4\mathcal{C}}-\mathcal{C}(N-N_*)^2\,, \qquad N_* = \frac{\mu}{2\mathcal{C}}\,.
 \label{qft:refinedranksaddle}
\end{equation}

We split the grand canonical partition function into perturbative and non-perturbative parts,
\begin{equation}
 \begin{split} \Xi^{\mathrm p}(\mu;x,y,u) &= \mathcal I_\infty\sum_{N\in\mathbb Z}e^{\mu N-\mathcal{C}N^2}\,,\quad  \Xi = \Xi^{\mathrm p}+\Xi^{\mathrm{np}}\,, \end{split}
 \label{qft:ensembleparts}
\end{equation}
and we sum over all integers to simplify the analysis. This extension gives additional non-perturbative corrections. The full non-perturbative grand canonical partition function equals
\begin{equation}
 \Xi^{\mathrm{np}}(\mu;x,y,u) =\sum_{N=0}^{\infty}e^{\mu N}Z_N^{\mathrm{np}} -\mathcal I_\infty\sum_{n=1}^{\infty}e^{-\mu n-\mathcal{C}n^2}\,.
 \label{qft:ensembleendpoint}
\end{equation}
Poisson summation evaluates the Gaussian sum in $\Xi^{\mathrm p}$ exactly,
\begin{equation}
 \begin{split} \Xi^{\mathrm p}(\mu;x,y,u) = \mathcal I_\infty\sqrt{\frac{\pi}{\mathcal{C}}} \sum_{\ell\in\mathbb Z}\exp\!\left[\frac{(\mu+2\pi\rmi\ell)^2}{4\mathcal{C}}\right] = \sum_{\ell\in\mathbb Z}e^{J^{\mathrm p}(\mu+2\pi\rmi\ell;x,y,u)}\,. \end{split}
 \label{qft:refinedrankasymptotics}
\end{equation}
where the perturbative grand potential is
\begin{equation}
 J^{\mathrm p}(\mu;x,y,u) = \frac{\mu^2}{4\mathcal{C}}+\log\mathcal I_\infty+\frac12\log\frac{\pi}{\mathcal{C}}\,,
 \label{qft:refinedmodified}
\end{equation}
and the factor $\sqrt{\pi/\mathcal{C}}$ comes from the Gaussian normalization of the ensemble transform.

\paragraph{The leading correction at general fugacities.}

We first keep all fugacities and derive the leading correction at fixed rank, \eqref{qft:leadingfixedrank}, and at fixed $\mu$, \eqref{qft:leadingrefined}. We can directly apply the analysis of~\cite{Murthy:2022ien,Liu:2022olj} for this. In the notation of \cite{Liu:2022olj} the determinant expansion, isolating the non-perturbative contributions, takes the form
\begin{equation}
    \frac{Z_N}{Z_{N}^p} = \frac{\rme^{\mathcal F_N} Z_N}{\mathcal I_{\infty}} = 1 + \sum_{m\geq 1} G^{(m)}_N\,.
\end{equation}
Writing $f_k=f_{\rm LS}(x^k,y^k,u^k)$ and $\gamma_k=f_k/(1-f_k)$, the first non-perturbative term is then 
\begin{equation}
G_N^{(1)} = -\oint \frac{\rmd \zeta}{2\pi \rmi} \zeta^{N-1} \frac{\zeta}{(1-\zeta)^2}
 \exp\!\left[-\sum_{k=1}^{\infty}\frac{\gamma_k}{k}(1-\zeta^k)(1-\zeta^{-k})\right]\,.
 \label{qft:firstgiantkernel}
\end{equation}
The poles contributing to this contour integral follow from expanding $\gamma_k$
\begin{equation}
    \gamma_k = (x u)^k + (x/u)^k + \mathcal O(x^{2k})\,.
\end{equation}
Explicitly evaluating the residues in \eqref{qft:firstgiantkernel}, at the poles $\zeta = (x u, x/u)$, the first non-perturbative corrections takes the form
\begin{equation}\label{qft:leadingfixedrank}
    G_N^{(1)} = \sum_{\sigma = \pm 1} \mathcal G_\sigma (x u^\sigma)^{N} + \mathcal O(x^{2N})\,,
\end{equation}
where
\begin{equation}
     \mathcal G_\sigma =  -\frac{(x^2;x^2)_\infty^2} {(x^2y;x^2)_\infty(x^2/y;x^2)_\infty} \frac{(xu^{-\sigma}y;xu^{-\sigma})_\infty (xu^{-\sigma}/y;xu^{-\sigma})_\infty} {\theta(xu^\sigma;xu^{-\sigma})} (x u^\sigma)\,.
\end{equation}

At $y=1$, we combine the the two sectors before taking $u\to1$. This recovers the leading unflavoured fixed-rank correction $Z_N^{\mathrm{np}}/Z_N^{\mathrm p}=-(N+2)x^{N+1}+\mathcal O(N^2x^{2N+4})$ for $N\to\infty$ at fixed $0<x<1$.

At fixed $\mu$, the non-perturbative corrections simply shift the chemical potential by $\rmi\pi(\tau+\sigma\varphi)$. Their weights relative to the perturbative contribution are therefore
\begin{equation}
 \begin{split}
 \mathcal W_\sigma(\mu)&=\exp\!\left[J^{\mathrm p}\bigl(\mu+\rmi\pi(\tau+\sigma\varphi)\bigr)-J^{\mathrm p}(\mu)\right] = \exp\!\left[\frac{\rmi\pi(\tau+\sigma\varphi)\mu}{2\mathcal C}-\frac{\pi^2(\tau+\sigma\varphi)^2}{4\mathcal C}\right]\,.
 \end{split}
 \label{qft:giantweights}
\end{equation}
Multiplying these weights by the residue coefficients $\mathcal G_\sigma$, which are independent of $\mu$, gives the first non-perturbative correction

\begin{equation}
 \begin{split}
 J^{\mathrm{np}}_{(1)}(\mu;x,y,u)&=\sum_{\sigma=\pm}\mathcal G_\sigma(x,y,u)\mathcal W_\sigma(\mu)\,,
 \end{split}
 \label{qft:leadingrefined}
\end{equation}
whose surpression action in terms of the variables $(\tau,\psi,\varphi)$ equals
\begin{equation}
    S_\sigma(\mu) = \frac{2(\tau^2 - \psi^2)}{\tau(\tau - \sigma \varphi)} \mu\,. \label{qft:leadingactions}
\end{equation}
At $u=1$ the two terms combine, and in particular at $y=u=1$ we find that
\begin{equation}
 J^{\mathrm{np}}_{(1)}(\mu;x,1,1) =-\frac{2\rmi\mu}{\pi\tau}e^{-2\mu}\,.
 \label{qft:leadingunrefined}
\end{equation}
This is a contribution to the modified potential, before summing the large gauge images. 

\paragraph{The full non-perturbative sector at $y=u=1$.}
We now specialize to $y=u=1$, or $\psi=\varphi=0$, where the gas is free and its spectrum can be resummed exactly. In this limit $d_k$ vanishes, so $A=V=0$. In this unrefined limit, we derive the modified grand potential \eqref{qft:modifiedpotential} and the grand canonical partition function \eqref{qft:thetaensemble}.

At $u=1$ the density matrix~\eqref{qft:kernel} has the Fourier representation
\begin{equation}
 \rho_{x,0}(\alpha,\alpha') =\frac1\pi\sum_{r\in\mathbb Z+\frac12} \frac{e^{2\rmi r(\alpha-\alpha')}}{x^r+x^{-r}}\,.
 \label{qft:fermispectrum}
\end{equation}
Summing over the fermion occupation numbers gives the Fredholm determinant, which takes the same form as for the Schur index studied in~\cite{Bourdier:2015wda} (see also~\cite{Bourdier:2015sga} for relevant generalisations),
\begin{equation}
 \det(1+\kappa\rho_{x,0}) =\prod_{r\in\mathbb Z+\frac12} \left(1+\frac{\kappa}{x^r+x^{-r}}\right)\,, \qquad \kappa=e^\mu\,.
 \label{qft:fredholm}
\end{equation}
To evaluate the determinant, we take its logarithm and rewrite it in the Schwinger proper-time representation. Using $\log(1+\kappa/a)=\int_0^\infty\dd t\,(e^{-at}-e^{-(a+\kappa)t})/t$ we find
\begin{equation}
 \begin{split} \log\det(1+e^\mu\rho_{x,0}) = \int_0^\infty\frac{\dd t}{t} (1-e^{-te^\mu})\,\mathcal K_x(t)\,,\qquad \mathcal K_x(t)&=\sum_{r\in\mathbb Z+\frac12} e^{-t(x^r+x^{-r})}\,. \end{split}
 \label{qft:schwinger}
\end{equation}
This is the logarithm of the free-gas determinant; the physical grand canonical partition function also keeps the normalization factors in~\eqref{qft:fermiprefactor}. To evaluate this intgral in a large-$\mu$ expansion, we rewrite it as a Mellin--Barnes contour integral\footnote{This Mellin--Barnes reparametrization was also shown to be particularly useful in the context of 3d CS theories in \cite{Marino:2016new}.}
\begin{equation}
 \begin{split}
 \log\det(1+e^\mu\rho_{x,0}) &=-\frac{\rmi}{\pi\tau} \sum_{m\in\mathbb Z}(-1)^m \int_{\mathcal{C}_{\rm MB}}\frac{\dd z}{2\pi\rmi}\, \mathcal R_m(z;\mu)\,,\\
 \mathcal R_m(z;\mu) &=\Gamma(-2z)\Gamma(z+\frac{m}{\tau})\Gamma(z-\frac{m}{\tau})e^{2z\mu}\,,
 \end{split}
 \label{qft:mellincontour}
\end{equation}
where the integral goes from $c-\rmi\infty$ to $c+\rmi\infty$, with $0<c<\tfrac12$. The kernel has three types of poles. There are simple poles at all positive half integers, coming from $\Gamma(-2z)$, a simple pole at the origin, coming from the same $\Gamma$-function and simple poles at $z = \pm \frac{m}{\tau}$. This contour integral determines the full non-perturbative grand potential.

\begin{figure}[htbp]
 \centering
\begin{tikzpicture}[x=1.65cm,y=1cm,>=Stealth,font=\small]
 \colorlet{contourblue}{blue!65!black}
 \colorlet{originorange}{orange!80!black}
 \colorlet{negativepurple}{violet!80!black}
 \tikzset{
   contour/.style={draw=contourblue,line width=0.9pt},
   closure/.style={contour,dashed},
 }

 \draw[->,black!45] (-3.8,0)--(2.45,0) node[right,text=black] {$\operatorname{Re}z$};
 \draw[->,black!35] (0,-2.85)--(0,2.9) node[above,text=black] {$\operatorname{Im}z$};

 \draw[contour] (.25,-2.2)--(.25,2.2);
 \draw[contour,->] (.25,-.48)--(.25,.48);
 \draw[closure] (.25,2.2)--(-3.25,2.2)--(-3.25,-2.2)--(.25,-2.2);
 \draw[contour,->] (-1,2.2)--(-1.65,2.2);
 \draw[contour,->] (-3.25,.35)--(-3.25,-.35);
 \draw[contour,->] (-1.65,-2.2)--(-1,-2.2);
 \node[contourblue,anchor=west] at (.34,.3) {$\mathcal{C}_{\rm MB}$};
 \node[black!60,anchor=south east] at (-.08,2.23) {$+\rmi T$};
 \node[black!60,anchor=north east] at (-.08,-2.23) {$-\rmi T$};

 \foreach \p/\lab in {-3/{-M},-2/{-2},-1/{-1}}{
   \fill[negativepurple] (\p,0) circle (2.6pt);
   \draw[negativepurple] (\p,0) circle (4pt);
   \node[below=6pt] at (\p,0) {$\lab$};
 }
 \node[black!60] at (-2.5,.05) {$\cdots$};

 \fill[originorange] (0,0) circle (2.6pt);
 \draw[originorange] (0,0) circle (4pt);
 \draw[originorange] (0,0) circle (5.4pt);
 \node[below=7pt] at (0,0) {$0$};

 \foreach \p/\lab in {.5/{1/2},1/{1},1.5/{3/2},2/{2}}{
   \fill[black!45] (\p,0) circle (2.2pt);
   \node[below=5pt,font=\footnotesize] at (\p,0) {$\lab$};
 }

 \fill[originorange] (-3.05,-3.25) circle (2.2pt);
 \draw[originorange] (-3.05,-3.25) circle (3.6pt);
 \draw[originorange] (-3.05,-3.25) circle (5pt);
 \node[anchor=west] at (-2.9,-3.25) {$m=0$: triple pole};
 \fill[negativepurple] (-.15,-3.25) circle (2.2pt);
 \draw[negativepurple] (-.15,-3.25) circle (3.6pt);
 \node[anchor=west] at (0,-3.25) {$m=0$: double poles};
\end{tikzpicture}
 \caption{The Mellin--Barnes contour and the pole locations of the zero-mode kernel $\mathcal R_0$. The upward line $\mathcal{C}_{\rm MB}$, with $0<c<\tfrac12$, closes counterclockwise to the left. The origin is a triple pole and the negative-integer poles are double; the positive simple poles remain outside the contour. First take $T\to\infty$, then $M\to\infty$, in the convergence domain.}
 \label{fig:mellincontour}
\end{figure}
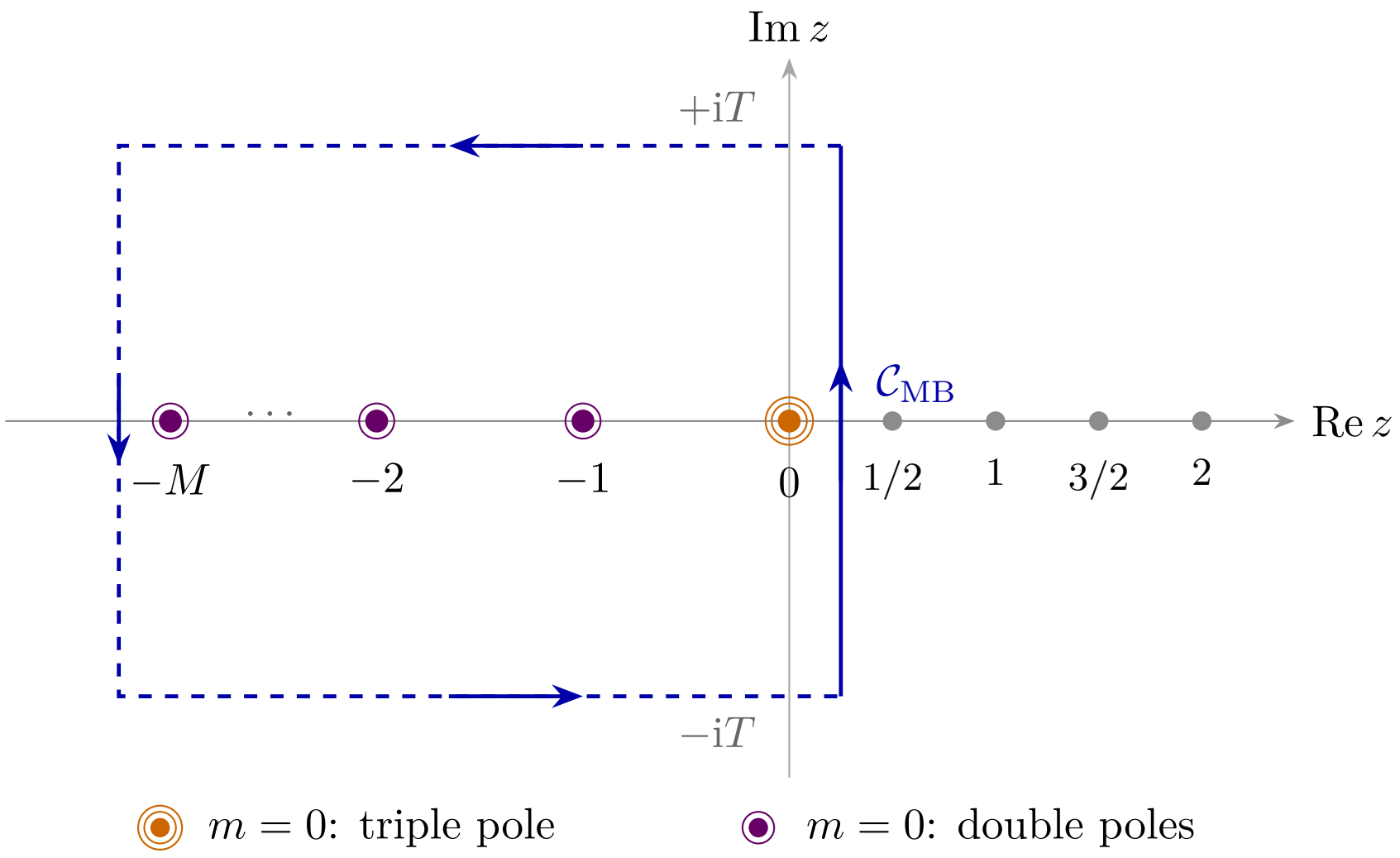

We first focus on the modified grand potential, which keeps the term $m=0$. The remaining poles come from making the chemical potential $\mu$ periodic along the imaginary axis. For $m=0$ the the kernel reduces to $\mathcal R_0(z;\mu)=\Gamma(-2z)\Gamma(z)^2e^{2z\mu}$, and the modified grand potential equals
\begin{equation}
 \mathcal J = -\frac{\rmi}{\pi\tau} \int_{\mathcal{C}_{\rm MB}}\frac{\dd z}{2\pi\rmi}\, \mathcal R_0(z;\mu)\,,
\end{equation}
which has simple poles on positive half-integers, a triple pole at the origin, and double poles at negative integers. To evaluate the integral, we close the contour as in figure \ref{fig:mellincontour}. The contributions of the horizontal lines vanish, and the contribution on the left vanishes at large $M$ for $\operatorname{Re}\mu>\log2$ and $|\operatorname{Im}\mu|<\pi$. The integral thus reduces to a sum of residues. The pole at the origin (the zero-pole) gives the perturbative contribution
\begin{equation}
 \mathcal J^p(\mu) = -\frac{\rmi}{\pi\tau} \operatorname*{Res}_{z=0}\mathcal R_0(z;\mu) = \frac{\rmi \mu^2}{\pi \tau} + \frac{\rmi\pi}{4\tau}\,,
\end{equation}
which we attribute to the supergravity degrees of freedom, as we will show in section \ref{sec:supergravity}.

The negative poles give the non-perturbatively surpressed contributions
\begin{equation}
 \mathcal J^{\mathrm{np}}(\mu) = -\frac{\rmi}{\pi\tau} \sum_{n=1}^{\infty}\operatorname*{Res}_{z=-n}\mathcal R_0(z;\mu) = \frac{\rmi}{\pi\tau} \sum_{n=1}^{\infty}\frac1n\binom{2n}{n} \bigl(H_{2n-1}-H_n-\mu\bigr)e^{-2n\mu}\,,
 \label{qft:unflavouredwinding}
\end{equation}
where $H_n = \sum_{j=1}^n j^{-1}$ are the harmonic numbers. These exponentially suppressed contributions provide targets for D3-brane sectors in the bulk. Section~\ref{sec:dbranes} derives the corresponding brane saddle and fluctuations in bulk variables at $y=1$. The leading grand-canonical target is
\begin{equation}
 \mathcal J^{\text{np}}_{1}(\mu) = -\frac{2\rmi \mu }{\pi \tau} \rme^{-2\mu}\,,
\end{equation}
which agrees with~\eqref{qft:leadingunrefined}. Section~\ref{d3:ensembles} obtains the same expression directly from the quantised bulk probe at fixed electric potential, including its spectral-zeta vacuum factor. Combining the perturbative and non-perturbative contributions, we find that the modified grand potential equals
\begin{equation}
 \mathcal J(\mu) = \frac{\rmi \mu^2}{\pi \tau} + \frac{\rmi\pi}{4\tau} +  \frac{\rmi}{\pi\tau} \sum_{n=1}^{\infty}\frac1n\binom{2n}{n} \bigl(H_{2n-1}-H_n-\mu\bigr)e^{-2n\mu}\,.
\end{equation}
As noticed in \cite{Bourdier:2015wda}, for the $\mathcal N=4$ Schur index, this answer can be resummed in terms of an effective chemical potential. It renormalizes the bare chemical potential non-perturbatively and is defined by
\begin{equation}
 \rme^{\mu} = 2\cosh \mu_{\text{eff}}\,,
\end{equation}
such that the modified grand potential takes the simple form 
\begin{equation}
 \mathcal J(\mu) = \frac{\rmi}{\pi \tau}\left(\mu_\text{eff}^2 + \frac{\pi^2}{4}\right)\,.
 \label{qft:modifiedpotential}
\end{equation}
As already stated above, the terms with $m\neq 0$ transform the exponent of the modified grand potential into the full grand canonical partition function, which is explicitly periodic in $\mu_{\text{eff}}$. Their explicit contributions are
\begin{equation}
 \begin{aligned}
 -\frac{2\rmi}{\pi \tau} \sum_{m = 1}^{\infty} (-1)^m \operatorname*{Res}_{z=0}\mathcal R_m(z;\mu) =& - \rmi \sum_{m=1}^{\infty}\frac{(-1)^m}{m \sin \frac{m\pi}{\tau}}\,,\\
 -\frac{2\rmi}{\pi \tau}\sum_{m = 1}^{\infty} (-1)^m \sum_{\sigma = \pm}^{}\operatorname*{Res}_{z= \sigma \frac{m}{\tau}}\mathcal R_m(z;\mu) =&\, 2\rmi \sum_{m=1}^{\infty}\frac{(-1)^m \cosh \frac{2 m \mu}{\tau}}{m \sin \frac{2 m\pi}{\tau}}\,,\\
 -\frac{2\rmi}{\pi\tau} \sum_{n,m=1}^{\infty}(-1)^m \sum_{\sigma=\pm1} \operatorname*{Res}_{z=\sigma \frac{m}{\tau}-n}\mathcal R_m(z;\mu)=&-\frac{2\rmi}{\tau}  \sum_{n,m=1}^{\infty}\sum_{\sigma=\pm1}\frac{(-1)^m \sigma}{n\sin\frac{2\pi m}{\tau}} \binom{2n-1-\frac{2\sigma m}{\tau}}{n-1} e^{2(\frac{\sigma m}{\tau}-n)\mu}\,.
 \end{aligned}
 \label{qft:nonzeroleadingpoles}
\end{equation}
Adding it all together the Fredholm determinant evaluates to the following combination of theta-functions
\begin{equation}
     \det(1+e^\mu\rho_{x,0}) = \frac{\vartheta_3(\rmi\mu_{\rm eff},x) + \dfrac{\vartheta_2(0,x)}{\vartheta_3(0,x)}\vartheta_2(\rmi\mu_{\rm eff},x)}{\vartheta_4(0,x)}\,.
     \label{qft:fullfredholm}
\end{equation}
The two theta blocks contain the even and odd powers of $\kappa=e^\mu$, respectively. On this slice the Casimir factor cancels $x^{-N^2/4}$ in~\eqref{qft:fermiprefactor}, while $1/\Delta_N(x,0)$ equals one at even rank and $\vartheta_3(0,x)/\vartheta_2(0,x)$ at odd rank. Applying this normalization coefficient by coefficient removes the prefactor of the odd block. Additionally we will use that $\vartheta_3(\rmi\mu_{\rm eff},x)+\vartheta_2(\rmi\mu_{\rm eff},x)=\vartheta_3(\rmi\mu_{\rm eff}/2,x^{1/4})$ to rewrite the answer in a simpler form below. The overall prefactor removes the constant $\rmi\pi/(4\tau)$ and restores $\mathcal I_\infty\sqrt{\pi/\mathcal{C}}$. Thus\footnote{We use the branch $\mu_{\mathrm{eff}}(\mu)\sim\mu$ at large positive $\mu$.}
\begin{equation}
 \begin{aligned}
 \Xi(\mu;x,1,1) &=\frac{\sqrt{4\rmi/\tau}\,e^{-\rmi\pi/(4\tau)}}{\vartheta_4(0,x)} \sum_{n\in\mathbb Z}e^{\mathcal J(\mu+2\pi\rmi n)} = \frac{\sqrt{4\rmi/\tau}}{\vartheta_4(0,x)} \sum_{n\in\mathbb Z} \exp\!\left[\frac{\rmi}{\pi\tau} \bigl(\mu_{\mathrm{eff}}(\mu)+2\pi\rmi n\bigr)^2\right]\,.
 \end{aligned}
 \label{qft:thetaensemble}
\end{equation}

\paragraph{Emergence from D-branes.} The separation between the pole at the origin and the nonzero real poles is reminiscent of emergence proposals for wrapped M2-branes~\cite{Hattab:2023moj,Hattab:2024thi,Hattab:2024ewk}, applied also in the holographic context in \cite{vanMuiden:2026nsp}. In these proposals, complexified Schwinger integrals encode both exponentially suppressed wrapped-brane contributions and polynomial terms in the supergravity prepotential. These integrals mirror the proper-time representation in \eqref{qft:mellincontour}, and the supergravity contributions are associated with the ultraviolet contribution at the origin.

We thus find an analogous structure in the protected LS partition function at fixed $\mu$\footnote{As a byproduct, we find that the same structure also holds in the grand canonical ensemble of the Schur background \cite{vanMuiden:2026lno}}. The triple pole of $\mathcal R_0$ at $z=0$ gives the quadratic term $\rmi\mu^2/(\pi\tau)=\mu^2/(4\mathcal C)=-S_\mu^{\mathrm{cl}}$, while the poles at $z=-n$ generate the $e^{-2n\mu}$ sectors in~\eqref{qft:unflavouredwinding}. Here $S_\mu^{\mathrm{cl}}$ is the fixed-potential action \eqref{sugra:grand-action}. The physical normalization in~\eqref{qft:thetaensemble} removes the additional constant $\rmi\pi/(4\tau)$ in the residue at the origin.

A single meromorphic kernel therefore encodes both the classical supergravity saddle and the exponential sectors whose leading term provides the target for D3-brane giant gravitons. This suggests a related emergence mechanism in an AdS setting with reduced supersymmetry and involving D-branes. Establishing it microscopically would require deriving the bulk effective action by integrating out the D3-brane degrees of freedom, beyond the comparison of the protected partition functions made here.

\section{Supergravity}\label{sec:supergravity}

The boundary partition functions are matched to bulk dynamics with appropriate boundary conditions on the magnetic flux and electric four-form period. In type IIB the relation between these boundary problems involves a topological term~\cite{Kurlyand:2022vzv,vanMuiden:2026lno,Adhikari:2026rfb,Adhikari:2026xdb}. We identify the two boundary problems from the four-form variation. We then evaluate the action in terms of the bulk radius, couplings and twist parameters, keeping the coefficient of the topological term explicit, before translating the result into the variables of each boundary ensemble.

\subsection{Background}\label{sugra:background}

The gravitational dual of the interacting LS fixed point is the Pilch--Warner solution of type IIB supergravity~\cite{Pilch:2000ej}. Its internal space is topologically $S^5$, but both its metric and fluxes differ from those of the maximally supersymmetric vacuum. We describe it using dimensionless complex embedding coordinates $(z_1,z_2,w)\in\mathbb C^3$, constrained by
\begin{equation}
 |z_1|^2+|z_2|^2+|w|^2=1\,.
 \label{sugra:embedding-constraint}
\end{equation}
We also introduce the quantities
\begin{equation}
 W=1+|w|^2\,,\qquad  k_z=\operatorname{Im}(\bar z_1\rmd z_1+\bar z_2\rmd z_2)\,,\qquad k_w=\operatorname{Im}(\bar w\rmd w)\,,
 \label{sugra:embedding-oneforms}
\end{equation}
in terms of which the ten-dimensional (Einstein-frame) metric is
\begin{equation}
 \begin{aligned} \rmd s_{10}^2&=L^2\sqrt W\left(\rmd s_{\mathrm{AdS}_5}^2+\rmd\tilde s_5^2\right)\,, \end{aligned}
 \label{sugra:embedding-metric}
\end{equation}
where the external and internal metrics are 
\begin{equation}
 \begin{aligned}
 \rmd s_{\mathrm{AdS}_5}^2& = \rmd\rho^2+\cosh^2\rho\,\rmd\hat\tau^2 + \sinh^2\rho\,\rmd\Omega_3^2\,,\quad \rmd\Omega_3^2=\rmd\eta^2+\sin^2\eta\,\rmd\chi_1^2 + \cos^2\eta\,\rmd\chi_2^2\,,\\
 \rmd\tilde s_5^2&=\frac4{3W} \left(|\rmd z_1|^2+|\rmd z_2|^2+\frac12|\rmd w|^2\right) + \frac4{9W^2}(k_z-k_w)^2\,.
 \end{aligned}
\end{equation}
The axion vanishes and the dilaton is constant, $C_0=0$ and $e^\Phi=g_s$. The two-form potentials take the form
\begin{equation}
 \begin{aligned} B_2-\rmi C_2&=\frac{2\rmi L^2}{3W}\sqrt{\frac23} \left[\rmd w\wedge(\bar z_1\rmd\bar z_2-\bar z_2\rmd\bar z_1) -2w\,\rmd\bar z_1\wedge\rmd\bar z_2\right]\,, \end{aligned}
 \label{sugra:fluxes}
\end{equation}
and the electric and magnetic four-form potentials are
\begin{equation}
 \begin{aligned} C_4^{\mathrm{el}}&=-\rmi L^4\sinh^4\rho\, \rmd\hat\tau\wedge\operatorname{vol}_{S^3}\,,\qquad C_4^{\mathrm{mag}}=-\frac{16L^4}{27W} \frac{k_w}{|w|^2}\wedge k_z\wedge\rmd k_z\,. \end{aligned}
 \label{sugra:fourform-potentials}
\end{equation}
Here $C_4=C_4^{\mathrm{el}}+C_4^{\mathrm{mag}}$, and $\operatorname{vol}_{S^3}$ is the volume form of the unit external three-sphere. The angular parametrization and explicit coordinate expressions for the background fields are given in appendix~\ref{app:backgroundcoordinates}.

The Page flux combines the five-form field strength with the two-form potentials to define a conserved, quantized charge~\cite{Marolf:2000cb}. To specify it, we first distinguish the closed curvature of $C_4$ from the gauge-invariant five-form:
\begin{equation}
 \begin{aligned}
 F_5& = F_5^{\mathrm{el}}+F_5^{\mathrm{mag}}\,, &F_5^{\mathrm{el}}&=\rmd C_4^{\mathrm{el}}\,, \quad F_5^{\mathrm{mag}} = \rmd C_4^{\mathrm{mag}}\,,\\
 \widetilde F_5&=F_5+X_5\,, &X_5&=\frac12(B_2\wedge F_3-C_2\wedge H_3)\,,
 \end{aligned}
 \label{sugra:polarization}
\end{equation}
where $H_3=\rmd B_2$ and $F_3=\rmd C_2$. The volume form of the unit round five-sphere, normalized by $\int_{S^5}\omega_5=\pi^3$, is
\begin{equation}
 \omega_5=\frac18(k_z+k_w)\wedge\rmd(k_z+k_w)\wedge\rmd(k_z+k_w)\,.
 \label{sugra:round-volume}
\end{equation}
The electric and magnetic components of the gauge-invariant five-form are then
\begin{equation}
 \widetilde F_5^{\mathrm{el}}=F_5^{\mathrm{el}} =-4\rmi L^4\operatorname{vol}_{\mathrm{AdS}_5}\,,\qquad \widetilde F_5^{\mathrm{mag}}=\frac{128L^4}{27W^2}\omega_5\,.
 \label{sugra:fiveform}
\end{equation}
Here $\operatorname{vol}_{\mathrm{AdS}_5}$ is the unit-radius volume form. We choose the ten-dimensional orientation with the internal volume form preceding the external one. With this orientation, the components obey
\begin{equation}
 \widetilde F_5^{\mathrm{el}}=-\rmi\star_{10}\widetilde F_5^{\mathrm{mag}}\,, \qquad \star_{10}\widetilde F_5=\rmi\widetilde F_5\,.
\end{equation}
This is the complex Euclidean continuation of the Lorentzian self-dual solution. 

On this background $B_2\wedge C_2=0$ and $X_5=B_2\wedge F_3$, so the Page flux reduces to
\begin{equation}
 \mathcal P_5=\widetilde F_5-B_2\wedge F_3 = \rmd\left(C_4-\frac12B_2\wedge C_2\right)=F_5\,,\qquad \int_{S^5}F_5^{\mathrm{mag}}=\frac{128\pi^3L^4}{27} \,.
 \label{sugra:page}
\end{equation}
Restoring the constant string coupling $g_s$ and writing $\ell_s$ for the string length, we write the D3-brane Page charge as
\begin{equation}
 N=\frac{1}{(2\pi \ell_s)^4g_s}\int_{S^5}F_5^{\mathrm{mag}}\,.
 \label{sugra:radius}
\end{equation}
The correspondingly normalized electric four-form period at
the boundary is
\begin{equation}
 \mu=\frac{\rmi}{(2\pi)^3g_s\ell_s^4} \left[\int_{S^1_\beta\times S^3}C_4^{\mathrm{el}}\right]_{\mathrm{ren}}\,.
 \label{sugra:chemical-potential}
\end{equation}
Here $S^1_\beta$ is the Euclidean time circle introduced in section~\ref{sugra:supersymmetry}. 

The definitions of $N$ and $\mu$ do not specify which is held fixed. The electric integral requires regularization and its conjugate interpretation and the ensemble boundary conditions are established in section~\ref{sugra:action}. 

\subsection{Supersymmetry and boundary fugacities}\label{sugra:supersymmetry}

To preserve supersymmetry we quotient the Euclidean background, making a twisted time direction periodic. To describe these twists explicitly, we first give an angular representation to the embedding coordinates:
\begin{equation}
 \begin{gathered}
 z_1=\cos\theta\cos\frac{\xi_1}{2}\,e^{\rmi(\xi_2+\xi_3)/2}\,,\qquad z_2=\rmi\cos\theta\sin\frac{\xi_1}{2}\,e^{\rmi(-\xi_2+\xi_3)/2}\,,\\
 w=\sin\theta\,e^{-\rmi\phi}\,,\qquad 0\leq\theta\leq\frac\pi2\,,\quad 0\leq\xi_1\leq\pi\,.
 \end{gathered}
 \label{sugra:embedding-coordinates}
\end{equation}
The background admits two Killing vectors associated with the
R-symmetry and flavour symmetry, respectively:
\begin{equation}
 K_R=\partial_{\xi_3}-\partial_\phi\,,\qquad K_F=2\partial_{\xi_2}\,.
 \label{sugra:generators}
\end{equation}
With respect to these $\text{U}(1)$ isometries, the embedding coordinates $(z_1,z_2,w)$ carry R-charges $(1/2,1/2,1)$ and flavour charges $(1,-1,0)$. For the external three-sphere, we use the angles $\eta,\chi_1,\chi_2$ of section \ref{sugra:background}, with $\chi_1,\chi_2$ of period $2\pi$. The two plane rotations are generated by the Killing vectors 
\begin{equation}
 K_a=\partial_{\chi_a}\,,\qquad J_1=j_1+j_2\,,\qquad J_2=j_2-j_1\,.
 \label{sugra:angular-generators}
\end{equation}
The supercharge selected in~\eqref{qft:supercharge} then has $(D,J_1,J_2,R,F)=(1/2,-1/2,-1/2,1,0)$.

Along these four isometry directions by $K_1, K_2, K_R, K_F$, we introduce angular twists $\epsilon_1,\epsilon_2$ and internal twists $\nu_R,\nu_F$. The periodicity then follows from the redefined twisted coordinates
\begin{equation}
 \begin{aligned}
 \chi_a&=\widetilde\chi_a+\rmi\epsilon_a\hat\tau\,, &\phi&=\widetilde\phi+\rmi\nu_R\hat\tau\,,\\
 \xi_3&=\widetilde\xi_3-\rmi\nu_R\hat\tau\,, &\xi_2&=\widetilde\xi_2-2\rmi\nu_F\hat\tau\,.
 \end{aligned}
 \label{sugra:periodic-coordinates}
\end{equation}
The tilded angles keep their usual angular periods, and we impose $\hat\tau\sim\hat\tau+\beta$. By~\eqref{sugra:embedding-coordinates}, the internal embedding coordinates transform as
\begin{equation}
 z_1\longmapsto e^{\beta(\nu_R/2+\nu_F)}z_1\,,\qquad z_2\longmapsto e^{\beta(\nu_R/2-\nu_F)}z_2\,,\qquad w\longmapsto e^{\beta\nu_R}w\,,
 \label{sugra:embedding-quotient}
\end{equation}
under this identification. The vector $Y$ generating this circle and the associated Hamiltonian $H_{\mathrm{twist}}$ are
\begin{equation}
 \begin{aligned}
 Y&=\partial_{\hat\tau}+\rmi\epsilon_1K_1+\rmi\epsilon_2K_2 -\rmi\nu_RK_R-\rmi\nu_FK_F\,,\\
 H_{\mathrm{twist}}&=D+\epsilon_1J_1+\epsilon_2J_2-\nu_RR-\nu_FF\,.
 \end{aligned}
 \label{sugra:twisted-hamiltonian}
\end{equation}
For the quotient to preserve supersymmetry, the chosen supercharge must commute with the twisted Hamiltonian. Using the charges of the supercharge as given above, this condition reads
\begin{equation}
 [H_{\mathrm{twist}},Q]=0\quad \Rightarrow \quad \frac12-\frac{\epsilon_1+\epsilon_2}{2}-\nu_R=0\,, \qquad \nu_R=\frac{1-\epsilon_1-\epsilon_2}{2}\,.
 \label{sugra:twist-supersymmetry}
\end{equation}
The flavour twist $\nu_F$ remains unconstrained because the Killing spinors are flavour-neutral. An explicit ten-dimensional Killing spinor and the corresponding projection conditions are given in appendix~\ref{app:explicitspinor}, where~\eqref{sugra:twist-supersymmetry} follows by differentiating the spinor along $Y$.

To connect these twist parameters with the boundary fugacities and complex moduli, we introduce the complex coordinates $\zeta_1=e^{-\hat\tau+\rmi\chi_1}\sin\eta$ and $\zeta_2=e^{-\hat\tau+\rmi\chi_2}\cos\eta$. Shifting $\hat\tau$ by $\beta$ at fixed tilded angles in~\eqref{sugra:periodic-coordinates} gives
\begin{equation}
 (\zeta_1,\zeta_2)\sim \left(e^{-\beta b_1}\zeta_1,e^{-\beta b_2}\zeta_2\right)\,, \qquad b_1=1+\epsilon_1\,,\qquad b_2=1+\epsilon_2\,.
 \label{sugra:hopf-twists}
\end{equation}
We can therefore identify the twists with the paramters in section \ref{sec:qft} through the map
\begin{equation}
 x=e^{-\beta(b_1+b_2)/4}\,,\qquad y=e^{\beta(b_2-b_1)/2}\,,\qquad u=e^{\beta\nu_F}\,, \qquad \nu_R=\frac{3-b_1-b_2}{2}\,.
 \label{sugra:fugacity-twists}
\end{equation}

For the round unflavoured twist we set $\epsilon_1=\epsilon_2=\nu_F=0$. Supersymmetry then gives $\nu_R=1/2$, while $b_1=b_2=1$ and $y=u=1$. The coordinate identifications become
\begin{equation}
 (\hat\tau,\phi,\xi_3)\sim (\hat\tau+\beta,\phi+\rmi\beta/2,\xi_3-\rmi\beta/2)\,, \qquad Y=\partial_{\hat\tau}-\frac{\rmi}{2}K_R\,,
 \label{sugra:quotient}
\end{equation}
which is the simplified twist that we focus on for the quantisation of D3-branes in section \ref{sec:dbranes}.

\subsection{Renormalized on-shell action and ensembles}
\label{sugra:action}

In this subsection, we compute the renormalized on-shell action at fixed flux and at fixed chemical potential. The two boundary problems differ in the coefficient $\alpha$ of a topological term: $\alpha =1$ at fixed flux and $\alpha =-1$ at fixed chemical potential. In the supersymmetric scheme, the fixed-flux action \eqref{sugra:canonical-action} equals $\mathcal{C}(x, y, u)N^2$
, where $\mathcal{C}(x, y, u)$ is the field-theory Casimir coefficient \eqref{qft:flavouredcasimir}. A Legendre transform of this action yields the fixed-potential action \eqref{sugra:grand-action}. 

We use the Pasti--Sorokin--Tonin (PST) formulation of type IIB supergravity~\cite{Pasti:1996vs,DallAgata:1997gnw}, in which five-form self-duality follows from the equations of motion together with the PST gauge symmetries. The relevant contributions in the Leigh--Strassler background are
\begin{equation}
 \begin{aligned}
 S_{\mathrm{PST}} ={}&-\frac{1}{2\kappa_{10}^2}\int_{M_{10}} \biggl[R_{10}\star_{10}1 -\frac12H_3\wedge\star_{10}H_3 -\frac12F_3\wedge\star_{10}F_3 -\frac14\widetilde F_5\wedge\star_{10}\widetilde F_5\biggr]\\
 &+\frac{\rmi}{4\kappa_{10}^2}\int_{M_{10}}C_4\wedge H_3\wedge F_3 +\frac{1}{8\kappa_{10}^2}\int_{M_{10}} \iota_v\mathcal F_5\wedge\star_{10}\iota_v\mathcal F_5\,,
 \end{aligned}
 \label{sugra:pst}
\end{equation}
where $\kappa_{10}^2=64\pi^7g_s^2\ell_s^8$, the five-form $\mathcal F_5=\widetilde F_5+\rmi\star_{10}\widetilde F_5$ is automatically self-dual and the vector $v_M=\partial_M a/\sqrt{-|\partial a|^2}$ is constructed from the auxiliary PST scalar $a$. For the solution given in section \ref{sec:supergravity}, we split the four-form into electric and magnetic parts, as in \eqref{sugra:fourform-potentials}. We keep this split even though the twists that we impose in \eqref{sugra:periodic-coordinates} mix some of the internal and external coordinates. A well-defined variational principle further requires appropriate boundary conditions, the Gibbons--Hawking--York term, and counterterms specified below. 

The boundary completion of the PST action involves two additional terms~\cite{Kurlyand:2022vzv,Adhikari:2026rfb},
\begin{equation}
 \begin{aligned} S_{\mathrm{top}} = -\frac{\rmi}{4\kappa_{10}^2} \int_{M_{10}}F_5^{\mathrm{el}}\wedge F_5^{\mathrm{mag}}\,,\qquad  S_X = -\frac{\rmi}{4\kappa_{10}^2} \int_{M_{10}}\rmd(C_4\wedge X_5)\,. \end{aligned}
 \label{sugra:topological}
\end{equation}
The inclusion of $S_X$ is necessary  to preserve the PST shift 
\begin{equation}
 \delta_b C_4=b_4\,,\qquad \delta_b a=0\,,\qquad \rmd a\wedge\rmd b_4=0\,,
 \label{sugra:pst-shift}
\end{equation}
with all other fields unchanged, so that $\delta_b X_5=0$. 

Under this transformation the PST action changes by a boundary term, while the variation of $S_X$ follows directly from its definition:
\begin{equation}
 \begin{aligned} \delta_b S_{\mathrm{PST}} = \frac{\rmi}{4\kappa_{10}^2} \int_{\partial M_{10}}b_4\wedge(F_5+X_5)\,,\quad \delta_b S_X = -\frac{\rmi}{4\kappa_{10}^2} \int_{\partial M_{10}}b_4\wedge X_5\,. \end{aligned}
 \label{sugra:pst-x-variation}
\end{equation}
Thus $S_X$ cancels the two-form-dependent part, while $S_{\mathrm{top}}$ cancels the remaining $b_4\wedge F_5$ term. 
Combining the three variations therefore gives~\cite{Kurlyand:2022vzv,Adhikari:2026rfb}
\begin{equation}
 \delta_b\left(S_{\mathrm{PST}}+S_X+S_{\mathrm{top}}\right) =\frac{\rmi}{4\kappa_{10}^2}\int_{\partial M_{10}} b_4\wedge\bigl[(F_5+X_5)-X_5-F_5\bigr]=0\,.
 \label{sugra:pst-boundary-cancellation}
\end{equation}
This cancellation holds for the unit coefficient of $S_{\mathrm{top}}$. 

To treat both boundary problems with one action, we give the topological term a coupling $\alpha$. We also add the Gibbons--Hawking--York term, which ensures that the boundary metric is held fixed, and a boundary term $S_{\mathrm{ct}}^{(\alpha)}$ renormalizing the chosen boundary problem. The total action is
\begin{equation}
 S^{(\alpha)}=S_{\mathrm{PST}}+S_X+\alpha S_{\mathrm{top}} +S_{\mathrm{GHY}}+S_{\mathrm{ct}}^{(\alpha)}\,.
 \label{sugra:ensemble-action}
\end{equation}
Its four-form sector has the residual PST variation
\begin{equation}
 \delta_b\left(S_{\mathrm{PST}}+S_X+\alpha S_{\mathrm{top}}\right) =\frac{\rmi(1-\alpha)}{4\kappa_{10}^2} \int_{\partial M_{10}}b_4\wedge F_5\,.
 \label{sugra:pst-alpha-variation}
\end{equation}

For the fixed-flux, or Neumann, boundary problem, the magnetic Page charge $N$ is held fixed while the electric period $\mu$ may vary. The PST shifts are then allowed to act at the boundary, and~\eqref{sugra:pst-alpha-variation} fixes $\alpha=+1$. The same value makes the action stationary under arbitrary variations of the electric period.

For Dirichlet boundary conditions, instead, the renormalized electric source is held fixed and the allowed PST shifts obey $b_4|_{\partial}=0$. The source subtraction is specified in Appendix~\ref{app:casimir:electric-potential}. The action is then stationary and its boundary PST variation vanishes for any $\alpha$, so at first sight its value appears unconstrained. With all other boundary sources held fixed, the variation of the action under a change of the electric potential, in a fixed flux sector, is~\cite{vanMuiden:2026lno}
\begin{equation}
 \left.\delta S^{(\alpha)}\right|_{\delta N=0} =-\frac{1-\alpha}{2}N\,\delta\mu\,.
 \label{sugra:ensemble-variation}
\end{equation}
To fix the consistent value of $\alpha$ for Dirichlet boundary conditions, we impose the standard thermodynamic normalization of the response, $\partial_\mu S_\mu=-N$. With \eqref{sugra:ensemble-variation}, this gives $\alpha = -1$. 

We now evaluate the on-shell action for both ensembles. If we impose the traced Einstein equations, the bulk action reduces entirely to a boundary term
\begin{equation}
 \begin{aligned} S_{\mathrm{PST}}\big|_{\mathrm{LS}} &= \frac{1}{8\kappa_{10}^2}\int_{\partial M_{10}} K_9 + \frac{\rmi}{4\kappa_{10}^2}\int_{\partial M_{10}} C_4\wedge X_5\,, \end{aligned}
 \label{sugra:pst-on-shell}
\end{equation}
where $K_9 = B_2\wedge\star_{10}H_3+C_2\wedge\star_{10}F_3$. The $K_9$ contribution vanishes on the LS background as $F_3$ and $H_3$ are entirely on the internal space. The contributions $\alpha S_{\mathrm{top}}$ and $S_X$ evaluate to
\begin{equation}
 \alpha S_{\mathrm{top}}=-\frac{\rmi \alpha}{4\kappa_{10}^2} \int_{\partial M_{10}}C_4^{\mathrm{el}}\wedge F_5^{\mathrm{mag}}\,,\qquad S_X = -\frac{\rmi}{4\kappa_{10}^2} \int_{\partial M_{10}} C_4\wedge X_5 \,.
 \label{sugra:boundary-pairing}
\end{equation}
The $S_X$ contribution cancels the remaining PST contribution, so we are left with $\alpha S_\text{top}$ and the boundary terms $S_{\text{GHY}}$ and $S_{\text{ct}}^{(\alpha)}$. We include these boundary contributions in a renormalized volume, whose prescription we leave unspecified for now. On the LS background the renormalized action is then
\begin{equation}
 \begin{aligned}
 S_{\mathrm{ren}}^{(\alpha)} = \left[-\frac{\alpha\rmi}{4\kappa_{10}^2} \int F_5^{\mathrm{el}}\wedge F_5^{\mathrm{mag}} + S_{\mathrm{GHY}}+S_{\mathrm{ct}}^{(\alpha)}\right]_{\mathrm{finite}} = \alpha\frac{2L^8}{27\pi^4g_s^2\ell_s^8}
 \operatorname{Vol}_{\mathrm{AdS}_5}^{\mathrm{ren}}\,.
 \end{aligned}
 \label{sugra:completed-bulk}
\end{equation}

As in~\cite{vanMuiden:2026lno}, we first evaluate the action in the fixed chemical potential ensemble, with $\alpha = -1$, and leave $\operatorname{Vol}_{\mathrm{AdS}_5}^{\mathrm{ren}}$ unspecified until choosing the renormalization prescription below. The background source, dual to the chemical potential~\eqref{sugra:chemical-potential}, is then
\begin{equation}
 \mu=\frac{2\pi \rmi}{(2\pi \ell_s)^4 g_s} \left[\int_{S^1_\beta\times S^3}C_4^{\mathrm{el}}\right]_{\mathrm{ren}} =\frac{L^4}{2\pi^3g_s\ell_s^4} \operatorname{Vol}_{\mathrm{AdS}_5}^{\mathrm{ren}}\,.
 \label{sugra:electric-period}
\end{equation}
Here the four-cycle lies at the asymptotic boundary and at fixed tilded internal coordinates. Substituting the renormalized potential back into the action gives
\begin{equation}
 S_{\mathrm{ren}}^{(\mu)} = - \frac{2L^8}{27\pi^4g_s^2\ell_s^8} \operatorname{Vol}_{\mathrm{AdS}_5}^{\mathrm{ren}} = - \frac{8\pi^2\mu^2}{27\operatorname{Vol}_{\mathrm{AdS}_5}^{\mathrm{ren}}}\,.
 \label{sugra:period-action}
\end{equation}

To explicitly evaluate the volume, we first use minimal subtraction. The Gibbons--Hawking--York term and the standard local counterterms $S_{\mathrm{ct,min}}$ remove the divergences and give
\begin{equation}
 \begin{gathered}
 \operatorname{Vol}_{\mathrm{AdS}_5}^{\mathrm{ren,min}} =\frac{3\pi^2\beta}{16}\,,\qquad
 \left.\mu\right|_{\mathrm{min}}=\frac{3\beta L^4}{32\pi g_s\ell_s^4}\,,\qquad  \left.S_{\mathrm{ren}}^{(\mu)}\right|_{\mathrm{min}}  = -\frac{128\mu^2}{81\beta}\,.
 \end{gathered}
 \label{sugra:minimal-volume}
\end{equation}
At fixed $\mu$ and $\beta$, this answer is independent of the angular and flavour twists, since the local geometric counterterms are insensitive to~\eqref{sugra:periodic-coordinates}. It therefore misses the fugacity dependence of the field-theory Casimir coefficient $\mathcal{C}(x,y,u)$ in~\eqref{qft:flavouredcasimir}. The leading large-$\mu$ contribution in the fixed-potential ensemble is $-\mu^2/[4\mathcal{C}(x,y,u)]$, as follows from~\eqref{qft:refinedmodified}. This mismatch arises because that the minimal subtraction is not the supersymmetric scheme, as was discussed in a similar setup in \cite{BenettiGenolini:2016tsn}.

We start with zero flavour twist ($\nu_F = 0$). In this case, the LS vacuum lies in a consistent truncation to minimal gauged supergravity, so we can use the supersymmetric holographic renormalization result of~\cite{BenettiGenolini:2016tsn}. We denote the volume including the supersymmetric boundary terms by $\operatorname{Vol}_{\mathrm{AdS}_5}^{\mathrm{SUSY}}$, with its precise definition given below. The volume, electric potential and on-shell action evaluate to
\begin{equation}
 \begin{gathered}
\operatorname{Vol}_{\mathrm{AdS}_5}^{\mathrm{SUSY}} = \frac{\pi^2\beta(b_1+b_2)^3}{54b_1b_2}\,,\quad \mu = \frac{\beta L^4(b_1+b_2)^3}{108\pi g_s\ell_s^4b_1b_2}\,,\quad S_{\mathrm{ren}}^{(\mu)} = -\frac{16b_1b_2\mu^2}{\beta(b_1+b_2)^3}\,.
 \end{gathered}
 \label{sugra:five-dimensional-casimir}
\end{equation}

To keep the flavour fugacity $u=e^{\beta\nu_F}$ non-trivial, we incorporate the contribution of a vector multiplet in the five-dimensional analysis of~\cite{BenettiGenolini:2016tsn}. Appendix~\ref{app:casimir} evaluates the supersymmetric finite boundary contribution, with its coefficient fixed by the bulk five-dimensional Chern--Simons coupling. The flat flavour connection leaves the local AdS geometry unchanged, but gives a finite boundary contribution. Including it in the supersymmetric subtraction, we define the common bulk coefficient $\mathcal{C}$ 
\begin{equation}
 \begin{gathered}
 \mathcal{C} = \frac{\beta(b_1+b_2)\bigl[(b_1+b_2)^2-16\nu_F^2\bigr]}{64b_1b_2}\,,
 \end{gathered}
 \label{sugra:flavoured-coefficient}
\end{equation}
in terms of which the supersymmetrically regularized volume, the chemical potential, and the on-shell action evaluate to
\begin{equation}
	\operatorname{Vol}_{\mathrm{AdS}_5}^{\mathrm{SUSY}} = \frac{32\pi^2}{27}\mathcal{C}\,,\qquad
  \mu = \frac{16L^4}{27\pi g_s\ell_s^4}\mathcal{C}\,,\qquad
 S_{\mathrm{ren}}^{(\mu)} = -\frac{\mu^2}{4\mathcal{C}}\,.
  \label{sugra:flavoured-casimir}
\end{equation}
The supersymmetric scheme for these quantities adds the finite boundary contribution $\alpha\Delta S_{\mathrm{susy}}$, where
\begin{equation}
 \Delta S_{\mathrm{susy}} =\frac{64L^8}{729\pi^2g_s^2\ell_s^8}
 \left(\mathcal{C}-\frac{81\beta}{512}\right)\,,
 \label{sugra:finite-completion}
\end{equation}
and we define the supersymmetric volume in terms of this finite contribution as  
\begin{equation}
 \begin{aligned}
 \operatorname{Vol}_{\mathrm{AdS}_5}^{\mathrm{SUSY}}
 = \int_{\mathrm{AdS}_5}\rmd^5x\sqrt{g_{\mathrm{AdS}_5}} +\frac{27\pi^4g_s^2\ell_s^8}{2\alpha L^8} \left(S_{\mathrm{GHY}}+S_{\mathrm{ct}}^{(\alpha)}\right) = \operatorname{Vol}_{\mathrm{AdS}_5}^{\mathrm{ren,min}} +\frac{27\pi^4g_s^2\ell_s^8}{2L^8}\Delta S_{\mathrm{susy}}\,.
 \end{aligned}
 \label{sugra:renormalized-volume}
\end{equation}
Appendix~\ref{app:casimir} gives more details on the evaluation of this volume and the chemical potential, in the presence of this finite counterterm.

The identifications in~\eqref{sugra:fugacity-twists} directly equate the bulk coefficient $\mathcal{C}$ defined in~\eqref{sugra:flavoured-coefficient} with the field-theory Casimir coefficient defined in \eqref{qft:flavouredcasimir}
\begin{equation}
 \mathcal{C}=\mathcal{C}(x,y,u)=-\frac{\rmi\pi\tau(\tau^2-\varphi^2)}{4(\tau^2-\psi^2)}\,.
 \label{sugra:flavoured-source-comparison}
\end{equation}

For the fixed-flux boundary problem, we must impose $\alpha=+1$ and fix the flux sector
\begin{equation}
 N=\frac{1}{(2\pi\ell_s)^4g_s}\int_{S^5}F_5^{\mathrm{mag}}
 =\frac{8L^4}{27\pi g_s\ell_s^4}\,.
 \label{sugra:page-rank}
\end{equation}
Substituting this relation into~\eqref{sugra:completed-bulk} gives
\begin{equation}
 S_N^{\mathrm{cl}} =\frac{27N^2}{32\pi^2}\operatorname{Vol}_{\mathrm{AdS}_5}^{\mathrm{SUSY}}
 =\mathcal{C}(x,y,u)N^2\,,
 \label{sugra:canonical-action}
\end{equation}
where we evaluated the volume in the supersymmetric scheme, i.e. with~\eqref{sugra:flavoured-casimir}.

At fixed fugacities $\beta,b_1,b_2,\nu_F$, the electric potential~\eqref{sugra:electric-period} obeys $\mu=2\mathcal{C}N=\partial_N S_N^{\mathrm{cl}}$. The fixed-potential action is therefore the Legendre transform of~\eqref{sugra:canonical-action}, 
\begin{equation}
 \begin{aligned}
 S_\mu^{\mathrm{cl}}(\mu)
 &=\left[S_N^{\mathrm{cl}}(N)-\mu N\right]_{N=N_*}
 =-\mathcal{C}N_*^2 = -\frac{\mu^2}{4\mathcal{C}(x,y,u)}\,.
 \end{aligned}
 \label{sugra:grand-action}
\end{equation}
Conversely, $-\partial_\mu S_\mu^{\mathrm{cl}}=N_*$ and $S_N^{\mathrm{cl}}(N_*)=S_\mu^{\mathrm{cl}}(\mu)+\mu N_*$. The classical modified grand potential is thus $J_{\mathrm{cl}}=-S_\mu^{\mathrm{cl}}=\mu^2/(4\mathcal{C})$. These relations between the supersymmetric actions require the correctly renormalized electric potential and chemical potential.

The action~\eqref{sugra:grand-action} describes a single saddle and must still be completed under large gauge transformations of the electric four-form. As was explained for round AdS$_5 \times S^5$ background in~\cite{vanMuiden:2026lno}, the boundary electric period is defined only modulo these transformations, which identify $\mu\sim\mu+2\pi\rmi$. The fixed-potential partition function must therefore be periodic, and the classical contribution enters through a sum of exponentiated saddle actions,
in 
\begin{equation}
 \begin{aligned}
 \sum_{\ell\in\mathbb Z}e^{-S_\mu^{\mathrm{cl}}(\mu+2\pi\rmi\ell)} = \sum_{\ell\in\mathbb Z}\exp\!\left[\frac{(\mu+2\pi\rmi\ell)^2}{4\mathcal{C}}\right] = \sqrt{\frac{\mathcal{C}}{\pi}}\,\vartheta_3\!\left(\frac{\rmi\mu}{2},e^{-\mathcal{C}}\right)\,.
 \end{aligned}
 \label{sugra:periodic-completion}
\end{equation}
Thus $-S_\mu^{\mathrm{cl}}$ supplies the classical modified grand potential, while the logarithm of the periodic partition function includes the image sum. This is the bulk counterpart of~\eqref{qft:refinedrankasymptotics}. Thus we find that the zero-pole in figure \ref{fig:mellincontour} indeed corresponds to the two-derivative supergravity action.

The prefactor $\mathcal I_\infty\sqrt{\pi/\mathcal{C}}$ also requires the fluctuation determinant of the supergravity fields and the ensemble normalization, which we determine next. 

\subsection{The 1-loop supergravity contribution}\label{sugra:kk}

We now include the Kaluza--Klein fluctuations about the Pilch--Warner background of section~\ref{sugra:background}, with the supersymmetric quotient and source dictionary of section~\ref{sugra:supersymmetry}. Together with the classical saddle and the global fluctuation of the Page flux, the one-loop contribution gives \eqref{sugra:perturbative-result}, which matches the perturbative grand potential \eqref{qft:refinedmodified}. 

The propagating modes are graded by $H_{\mathrm{twist}}$ in~\eqref{sugra:twisted-hamiltonian}, so we keep the same angular and flavour fugacities as in the classical action. The eight short-multiplet towers are known from the exceptional-field-theory analysis of~\cite{Bobev:2020lsk}. To compare with the $\mathrm U(N)$ observable we also keep the centre contribution. The resulting single-particle index is
\begin{equation}
 i_{\mathrm{pert}}^{\mathrm U}(x,y,u) =\frac{x^2}{1-x^2}+\frac{xu}{1-xu}+\frac{x/u}{1-x/u} -\frac{x^2y}{1-x^2y}-\frac{x^2/y}{1-x^2/y}\,.
 \label{sugra:kk-character}
\end{equation}
Plethystic exponentiation gives
\begin{equation}
 \mathcal I_\infty(x,y,u)=\PE\!\left[i_{\mathrm{pert}}^{\mathrm U}(x,y,u)\right]
 =\frac{(x^2y;x^2y)_\infty(x^2/y;x^2/y)_\infty}
 {(x^2;x^2)_\infty(xu;xu)_\infty(x/u;x/u)_\infty}\,,
 \label{sugra:kk-product}
\end{equation}
matching the protected oscillator factor in~\eqref{qft:infinitefluctuations}.  

The one-loop determinant of these modes, computed in the supersymmetric scheme, introduces no additional Casimir factor. This can be verified directly from \eqref{sugra:kk-character} with a standard $\zeta$-function regularization. We define the frequencies
\begin{equation}
 \omega_0=\frac{b_1+b_2}{2}\,,\qquad
 \omega_\pm=\frac{b_1+b_2}{4}\mp\nu_F\,,\qquad
 x^2=e^{-\beta\omega_0}\,,\quad xu^{\pm1}=e^{-\beta\omega_\pm}\,,
 \label{sugra:kk-frequencies}
\end{equation}
with $x^2y=e^{-\beta b_1}$ and $x^2/y=e^{-\beta b_2}$. Each tower of in \eqref{sugra:kk-character} with frequency $\omega$ contributes $\omega^{-z} \zeta(z)$, such that 
\begin{equation}
 \begin{aligned}
 \zeta_{\mathrm{pert}}^{\mathrm U}(z)
 &=\left[\omega_0^{-z}+\omega_+^{-z}+\omega_-^{-z}-b_1^{-z}-b_2^{-z}\right]\zeta_{}(z)\,,\\
 E_{\mathrm{pert}}^{\mathrm U}
 &=\frac12\zeta_{\mathrm{pert}}^{\mathrm U}(-1)
 =-\frac{\omega_0+\omega_++\omega_--b_1-b_2}{24}=0\,.
 \end{aligned}
 \label{sugra:vacuum}
\end{equation}
Thus we find that \eqref{sugra:kk-product} is the finite part of the bulk one-loop partition function. The background does not allow for any normalizable zero-modes that could contribute a $\log \ell_s$ term, and empirically we find that there are no logarithmic contributions coming from UV-divergences. \footnote{The next local supergravity contribution is expected at order $(L/\ell_s)^2 \sim \mu^{1/2}$, corresponding to the eight derivative correction. We assume that such higher derivative terms vanish in this supersymmetric background.} 

Combining the classical saddle with the protected one-loop contribution gives the following terms in the perturbative type IIB expansion
\begin{equation}
 \mathcal Z_{\mathrm{IIB}}(\mu)
 =-S_\mu^{\mathrm{cl}}+\log\mathcal I_\infty
 =\frac{\mu^2}{4\mathcal C}+\log\mathcal I_\infty\,.
\end{equation}
Compared with the boundary result~\eqref{qft:refinedmodified}, this expression lacks the term $\frac12\log(\pi/\mathcal C)$. The global fluctuation of the Page flux supplies this $\mu$-independent term.

At fixed electric potential, the boundary condition allows the flux to fluctuate. Let $n$ denote the continuous momentum conjugate to the boundary source $\mu$, so that, as in \eqref{sugra:ensemble-variation}
\begin{equation}
	\left.\delta_\mu S\right|_{\partial}=-n\,\delta\mu\,.
\end{equation}
At fixed momentum the on-shell action and the response of the source evaluate to
\begin{equation}
	S_n^{\text{cl}} = \mathcal C n^2 \,,\quad \mu(n) = \partial_n S_n^{\text{cl}} = 2 \mathcal C n\,.
\end{equation}
Instead, the generating function of flux variations, at fixed source $\mu$, equals
\begin{equation}
 \begin{aligned}
 S_\mu^{} = \mathcal C n(\mu)^2-\mu \,n(\mu) = S_\mu^{\text{cl}} + \mathcal C (\delta n)^2\,,
 \end{aligned}
 \label{sugra:page-generating-function}
\end{equation}
where in the second equality we expand around the saddle point $n = n_0 + \delta n$ with $n_0 = \mu/(2\mathcal C)$. Integrating over the quadratic fluctuation gives the contribution
\begin{equation}
	\int \rmd (\delta n) \rme^{-\mathcal C (\delta n)^2} = \sqrt{\frac{\pi}{\mathcal C}}
\end{equation}
to the IIB partition function at fixed $\mu$, which becomes
\begin{equation}
	\mathcal Z_{\text{IIB}}^{\text{pert}}(\mu) = \frac{\mu^2}{4\mathcal C}+\log\mathcal I_\infty
 +\frac12\log\frac{\pi}{\mathcal C}\,,
 \label{sugra:perturbative-result}
\end{equation}
matching the perturbative grand potential in \eqref{qft:refinedmodified} exactly.

Including the large gauge images $\mu\mapsto\mu+2\pi\rmi\ell$, as in~\eqref{sugra:periodic-completion}, gives the full perturbative bulk partition function
\begin{equation}
 \begin{aligned}
 Z_{\mathrm{IIB}}^{\mathrm{pert}}
 &=\sum_{\ell\in\mathbb Z}
 \exp\!\left[\mathcal Z_{\mathrm{IIB}}^{\mathrm{pert}}(\mu+2\pi\rmi\ell)\right] = \mathcal I_\infty\, \vartheta_3\!\left(\frac{\rmi\mu}{2},e^{-\mathcal C}\right)\,.
 \end{aligned}
 \label{sugra:perturbative-periodic-completion}
\end{equation}

\section{D-branes}\label{sec:dbranes}

Wrapped D-branes provide non-perturbative saddles in the IIB partition function, in addition to the supergravity contributions discussed above. We quantize the leading non-perturbative D-brane saddle and compare its contribution with the leading non-perturbative correction in the grand canonical ensemble of
the field theory. In section \ref{d3:ensembles} we find that, at $y = 1$ and with the brane Casimir factor included,
the two match directly, before the sum over large gauge images.

We work in the  background with $b_1=b_2=1$, so that $y=1$, $x=e^{-\beta/2}$ and $\nu_R=1/2$. We keep the flavor fugacity $u=e^{\beta\nu_F}$ general, so only $x$ and $u$ are nontrivial. In this limit, $\mathcal C=\beta(1-4\nu_F^2)/8$.
 
\subsection{Embedding}\label{d3:embedding}

We find three supersymmetric brane embeddings. Two of them, the \textit{light} embeddings, lie at $z_1=0$ and $z_2=0$ respectively; the third, \textit{heavy} embedding, lies at $w=0$. We call the corresponding D3-branes the light giants and the heavy giant. The $\kappa$-symmetry analysis below shows that all embeddings localize to $\rho=0$. The heavy embedding has non-trivial fermionic zero-modes and does not contribute to the partition function, as shown in appendix~\ref{app:heavygiant}. We therefore focus on the light embeddings. 

We do the analysis at $z_1=0$ and comment on the $z_2$ embedding throughout. The flavour transformation $(z_1,z_2,w)\mapsto(z_2,-z_1,w)$ exchanges the two light embeddings and maps the twisted background at $\nu_F$ to that at $-\nu_F$. 

The light embedding at $z_1=0$ is
\begin{equation}
 \rho=0\,,\qquad z_1=0\,,\qquad
 z_2=e^{(1/4-\nu_F)\hat\tau}\cos\theta\,e^{\rmi\chi}\,,\qquad
 w=e^{\hat\tau/2}\sin\theta\,e^{-\rmi\phi}\,.
 \label{d3:coordinates}
\end{equation}
We define the pulled-back angular one-forms
\[
 \mathcal D\phi=\rmd\phi+\frac{\rmi}{2}\rmd\hat\tau\,,\qquad
 \mathcal D\chi=\rmd\chi+\rmi\left(\nu_F-\frac14\right)\rmd\hat\tau\,,
\]
in terms of which the induced metric on the twisted worldvolume is
\begin{equation}
 \frac{\rmd s_{{\rm D3}}^2}{L^2\sqrt W}
 =\rmd\hat\tau^2+\frac23\rmd\theta^2
 +\frac{2\sin^2\theta\cos^2\theta}{3W^2}(2\mathcal D\chi-\mathcal D\phi)^2
 +\frac{16}{9W^2}(\cos^2\theta\,\mathcal D\chi+\sin^2\theta\,\mathcal D\phi)^2\,.
 \label{d3:metric}
\end{equation}

\subsection{Kappa symmetry}\label{d3:kappa}

Before evaluating the on-shell action of the brane, we first check that the configuration preserves the background supersymmetry by performing a $\kappa$-symmetry analysis. For vanishing worldvolume flux, the analysis shows that supersymmetry requires $\rho=0$ and that the light giant is half-BPS. 

To derive this, we leave the radial position $\rho$ of the brane free, and let the worldvolume span $\hat\tau$ and the coordinates $(\theta,\phi,\chi)$. On the cycle $z_1=0$ we have $\xi_1=\pi$, so that $\sigma_1-\rmi\sigma_2$ vanishes while $\sigma_3=2\rmd\chi$, where the $\sigma_i$ are the left-invariant one-forms \eqref{sugra:forms}. The external one-forms $e^2,\ldots,e^5$ of~\eqref{app:spinor-external-frame} vanish on the worldvolume, as do $e^7$ and $e^8$. The nonvanishing pullbacks of the frames in~\eqref{app:spinor-external-frame}--\eqref{app:spinor-internal-frame} are
\begin{equation}
 \begin{aligned}
 P[e^1]&=LW^{1/4}\cosh\rho\,\rmd\hat\tau\,,& P[e^6]&=L\sqrt{\frac23}W^{1/4}\rmd\theta\,,\\
 P[e^9]&=L\sqrt{\frac23}\frac{\sin\theta\cos\theta}{W^{3/4}}(2\rmd\chi-\rmd\phi)\,,& P[e^{10}]&=\frac{4L}{3W^{3/4}}(\cos^2\theta\,\rmd\chi+\sin^2\theta\,\rmd\phi)\,.
 \end{aligned}
 \label{d3:pulled-back-frame}
\end{equation}

Throughout this section the $\Gamma$-matrices $\Gamma_a$ anti-commute and obey $\{\Gamma_a,\Gamma_b\}=2\delta_{ab}$, where the flat indices follow those of the frames in~\eqref{app:spinor-external-frame}--\eqref{app:spinor-internal-frame}. We define $\Gamma_*=-\rmi\Gamma_1\cdots\Gamma_{10}$. The IIB spinor is in a doublet $\epsilon=(\varepsilon,\varepsilon^c)^T$ which transforms under the Pauli matrices $s_i$. The pulled-back worldvolume matrices are
\begin{equation}
 \widehat\gamma_i=P[e^a]_i\,\Gamma_a\,,\qquad \{\widehat\gamma_i,\widehat\gamma_j\}=2h_{ij}\,.
 \label{d3:pulled-back-gammas}
\end{equation}
For vanishing worldvolume flux, the $\kappa$-projector simplifies to
\begin{equation}
 \begin{aligned}
 \Gamma_\kappa&=s_3\Gamma_{(4)}
 =\begin{pmatrix}\Gamma_{(4)}&0\\0&-\Gamma_{(4)}\end{pmatrix}\,,
 \qquad  \Gamma_{(4)}
 =-\Gamma_{69}\Gamma_{1\,10}\,,\qquad s_3=\operatorname{diag}(1,-1)\,,
 \end{aligned}
 \label{d3:kappa-projector}
\end{equation}
where $\Gamma_{(4)}^2=\Gamma_\kappa^2=1$. The projection condition on the ten-dimensional spinors is
\begin{equation} 
	\Gamma_\kappa\epsilon=\epsilon\,.
    \label{d3:kappaProj}
\end{equation}
This projection condition must hold at every point of the worldvolume. We impose it by commuting the projector through the Killing spinor~\eqref{app:coordinate-killing-spinor}, which can be written in a factorised form as
\begin{equation}
	\epsilon=\mathcal U_1\,\mathcal U_\rho\,\mathcal U_2\epsilon_0
	 \label{d3:kappa-equation}
\end{equation}
where
\begin{equation}
 \begin{gathered}
 \mathcal U_1=e^{-\rmi\pi s_3/4}\,W^{1/8}\,e^{\alpha(\theta)\mathcal N}\,,\quad
 \mathcal N=\tfrac12\bigl(\sqrt3\,\Gamma_{9\,10}+s_1\Gamma_{7\,10}\bigr)\,,\quad
 \mathcal U_\rho=e^{\frac{\rmi\rho}{2}s_3\Gamma_{5\,10}}\,,\\
 \mathcal U_2=e^{-\frac\eta2\Gamma_{25}}e^{-\frac{\chi_1}2\Gamma_{32}}e^{-\frac{\chi_2}2\Gamma_{45}}e^{\frac{\rmi\hat\tau}2s_3\Gamma_{1\,10}}e^{\rmi\phi s_3}\,,
 \end{gathered}
 \label{d3:spinor-factors}
\end{equation}
where $\alpha(\theta)$ is the angle~\eqref{app:spinor-angle} and the constant spinor $\epsilon_0$ obeys the background projections~\eqref{app:background-spinor-projections}. Almost all $\Gamma$-matrix bilinears commute with the $\kappa$-projector, except $s_3\Gamma_{5\,10}$, which anti-commutes, such that
\begin{equation}
 [\Gamma_\kappa,\mathcal U_1]=[\Gamma_\kappa,\mathcal U_2]=0\,,\qquad
 \Gamma_\kappa\,\mathcal U_\rho=\mathcal U_\rho^{-1}\,\Gamma_\kappa\,.
 \label{d3:projector-commutation}
\end{equation}
Inserting \eqref{d3:kappa-equation} into the projection condition and
multiplying by $\mathcal U_\rho\,\mathcal U_1^{-1}$, we find that the kappa condition is equivalent to
\begin{equation}
 \mathcal U_2\,\Gamma_\kappa\epsilon_0=\mathcal U_\rho^2\,\mathcal U_2\epsilon_0
 =\bigl(\cosh\rho+\rmi\sinh\rho\,s_3\Gamma_{5\,10}\bigr)\mathcal U_2\epsilon_0\,.
 \label{d3:kappa-reduced}
\end{equation}

To solve \eqref{d3:kappa-reduced}, we decompose $\epsilon_0 = \epsilon_0^+ + \epsilon_0^-$ into eigenspinors of $\rmi s_3\Gamma_{1,10}$. Equation~\eqref{d3:kappa-reduced} then reduces to 
\begin{equation}
 \begin{gathered}
 e^{\pm\hat\tau/2}\,\mathcal U_{\rm rot}\,e^{\rmi\phi s_3}(\Gamma_\kappa-\cosh\rho)\epsilon_0^\pm
 =\rmi\sinh\rho\;e^{\mp\hat\tau/2}\,s_3\Gamma_{5\,10}\,\mathcal U_{\rm rot}\,e^{\rmi\phi s_3}\epsilon_0^\mp\,,\\
 \mathcal U_2=\mathcal U_{\rm rot}\,e^{\frac{\rmi\hat\tau}2s_3\Gamma_{1\,10}}e^{\rmi\phi s_3}\,.
 \end{gathered}
 \label{d3:kappa-localisation}
\end{equation}
Both sides of this equation must vanish, which happens only for $\rho = 0$ and $\Gamma_{\kappa} \epsilon_0 = \epsilon_0$. With~\eqref{app:background-spinor-projections}, this condition on the constant spinor becomes
\begin{equation}
 \rmi\Gamma_{1\,10} \epsilon_0= \epsilon_0\,,
 \label{d3:kappa-constant-spinor}
\end{equation}
which halves the eight complex parameters left by~\eqref{app:background-spinor-projections} to four, so the light giant is half-BPS. 

\subsection{On-shell action}\label{d3:action}

We now compute the on-shell action of the two light giants. The result is \eqref{d3:classical-weight}. 
The effective D3-brane action, with the constant dilaton absorbed, equals
\begin{equation}
 \begin{aligned}
 S_{\text{D3}}&= \frac{2\pi}{(2\pi \ell_s)^4 g_s} \int\left(\rmd^4\sigma\,\sqrt{\det(P[g]+\mathcal F)}
             -\rmi\left(P[C]\wedge e^{\mathcal F}\right)_4\right),\quad \mathcal F = 2\pi\ell_s^2F_{}-P[B_2]\,.
 \end{aligned}
 \label{d3:DBIWZ}
\end{equation}
Here $C$ is the RR potential polyform, and $(\cdots)_4$ selects its four-form component. On the classical embedding, we have that $F_{\rm wv}=0$ and $P[B_2]=P[C_2]=P[C_4]=0$. The DBI part of the action then evaluates to
\begin{equation}
	S_{\text{DBI}} = \beta\Delta \,,\quad \Delta=\frac{2L^4}{9\pi g_s\ell_s^4}\, .
\end{equation}

The four-form in \eqref{sugra:fourform-potentials} is not globally well defined and is singular at at $w=0$, where the $\phi$ circle collapses. To evaluate the Wess--Zumino (WZ) contribution correctly, we perform a gauge transformation $C_4\to C_4+\rmd\Lambda_3$ such that the resulting pullback is regular. Keeping $\nu_R$ explicit for now, the $z_1=0$ cycle sits at the boundary of $\mathcal B_5^+\simeq S^1_\beta\times D^4$ at $\rho=0$, parametrized by
\begin{equation}
 \begin{aligned}
 z_1&=\cos\theta\cos\frac{\xi_1}{2}\,e^{(\nu_R/2+\nu_F)\hat\tau}\,,\quad z_2 =\cos\theta\sin\frac{\xi_1}{2}\,e^{(\nu_R/2-\nu_F)\hat\tau+\rmi\chi}\,,\quad w=\sin\theta\,e^{\nu_R\hat\tau-\rmi\phi}\,.
 \end{aligned}
 \label{d3:rr-filling}
\end{equation}
In terms of the twisted coordinates
\begin{equation}
	 \mathcal D\phi=\rmd\phi+\rmi\nu_R\rmd\hat\tau\,,\qquad \mathcal D\xi_2=\rmd\xi_2-2\rmi\nu_F\rmd\hat\tau\,,\qquad \mathcal D\xi_3=\rmd\xi_3-\rmi\nu_R\rmd\hat\tau\,,
\end{equation}
the magnetic potential in~\eqref{sugra:fourform-potentials} equals
\begin{equation}
	C_4^{\rm mag}=-\frac{4L^4\cos^4\theta}{27W}\sin\xi_1\, \mathcal D\phi\wedge\rmd\xi_1\wedge\mathcal D\xi_2\wedge\mathcal D\xi_3\,,
\end{equation} 
which vanishes at $\xi_1 = 0$ but is finite at $\theta = 0$. To resolve this, we add a pure gauge term
\begin{equation}
 \begin{aligned}
 \Lambda_3&=-\frac{4L^4\cos^4\theta}{27W}(1-\cos\xi_1)
       \mathcal D\phi\wedge\mathcal D\xi_2\wedge\mathcal D\xi_3\,,\\
 C_4^{\rm full}&=C_4^{\rm mag}+\rmd\Lambda_3 = \frac{8L^4\cos^3\theta\sin\theta(3+\sin^2\theta)}{27W^2}
       (1-\cos\xi_1)\,\rmd\theta\wedge\mathcal D\phi
       \wedge\mathcal D\xi_2\wedge\mathcal D\xi_3\,,
 \end{aligned}	
\end{equation}
which now vanishes at both $\xi_1 = 0$ and $\theta = 0$. The pull-back of this four-form onto the worldvolume is non-trivial
\begin{equation}
	 P[C_4^{\rm fill}]
 =-\frac{32\rmi L^4}{27}\left(\frac{\nu_R}{2}+\nu_F\right)
   \frac{\cos^3\theta\sin\theta(3+\sin^2\theta)}{W^2}
   \rmd\hat\tau\wedge\rmd\theta\wedge\rmd\phi\wedge\rmd\chi\,,
\end{equation}
and its contribution to the WZ action evaluates to
\begin{equation}
 \begin{aligned}
 S_{\mathrm{WZ}} = -\frac{8\beta L^4}{27\pi g_s\ell_s^4}
       \left(\frac{\nu_R}{2}+\nu_F\right)\,.
 \end{aligned}
 \label{d3:rr-holonomy}
\end{equation}
For the second light D3-giant, at $z_2 = 0$, the WZ contribution follows from \eqref{d3:rr-holonomy} with $\nu_F\mapsto-\nu_F$. 

The R- and flavor-charges of the brane are determined by the coefficients of the corresponding twist parameters, while its mass follows from the DBI action. These quantities are given by
\begin{equation}
 (\Delta,R,F)_\sigma=\frac{4L^4}{27\pi g_s\ell_s^4}
 \left(\frac32,1,2\sigma\right)\,.\qquad
 \sigma=+1\ (z_1=0)\,,\quad \sigma=-1\ (z_2=0)\,.
 \label{d3:charges}
\end{equation}
We find that $\Delta=3R/2$ as supersymmetry requires. Adding the DBI and WZ contributions and setting $\nu_R = 1/2$, to specialise to the $y=1$ background, we obtain 
\begin{equation}
 \begin{aligned}
 S_{\rm D3}^{\sigma}&=\beta(\Delta-R/2-\nu_FF)
 =\frac{4\beta L^4}{27\pi g_s\ell_s^4}(1-2\sigma\nu_F)\,.
 \end{aligned}
 \label{d3:classical-weight}
\end{equation}
The heavy giant $w=0$ brane has action $\frac{8\beta L^4}{27\pi g_s\ell_s^4}$, and its spectrum of fluctuations differs from that of the light giants. In appendix~\ref{app:heavygiant} we show that its one-loop contribution vanishes.

\subsection{One-loop quantisation}\label{d3:one-loop}

We now compute the Gaussian fluctuations around the light giant at $z_1=0$. The worldvolume theory contains six transverse scalars, a $\mathrm U(1)$ gauge field and the kappa-fixed D3 fermions. We keep $b_1=b_2=1$ and $u=e^{\beta\nu_F}$ throughout. The one-loop determinants of the two giants are related by $u\mapsto u^{-1}$.

\subsubsection{Bosonic fluctuations}

We obtain the quadratic action on the D3-brane worldvolume from the geometric quadratic action of~\cite{Gautason:2024nru}. For its bosonic part, we can rewrite it as the bosonic action of an abelian vector multiplet and a neutral hypermultiplet with non-trivial background couplings as in~\cite{Hama:2012bg}. The result is \eqref{d3:boson-action}. We take $\Phi=\Phi^i n_i$ to be the normal scalars on the brane, with $n_1,\ldots,n_4$ along the AdS directions and $n_5,n_6$ along the normal directions in the internal space. In terms of the pulled back fields the vector, and hyper multiplet constituents equal
\begin{equation}
 \begin{aligned}
 a&=\frac{1}{\sqrt{4\pi g_s}}\left(\delta A_{\rm wv}
       -\frac{P[\iota_\Phi B_2]}{2\pi\ell_s^2}\right)\,,\qquad F=\rmd a\,,\\
 \varphi&=-\frac{\sqrt\pi\,(\Phi^5-\rmi\Phi^6)}{2(2\pi\ell_s)^2\sqrt{g_s}}\,,\qquad
 \bar\varphi=\frac{\sqrt\pi\,(\Phi^5+\rmi\Phi^6)}{2(2\pi\ell_s)^2\sqrt{g_s}}\,,\\
 q_\alpha&=\frac{\sqrt\pi\,(\Phi^{2\alpha-1}+\rmi\Phi^{2\alpha})}{(2\pi\ell_s)^2\sqrt{g_s}}\,,\qquad
 \bar q_\alpha=\frac{\sqrt\pi\,(\Phi^{2\alpha-1}-\rmi\Phi^{2\alpha})}{(2\pi\ell_s)^2\sqrt{g_s}}\,,\quad \alpha=1,2\,.
 \end{aligned}
 \label{d3:proper-fluctuations}
\end{equation}
The fields $(a,\varphi,\bar\varphi)$ belong to the vector multiplet, while $q_1,q_2$ are the two complex scalars of the neutral hypermultiplet. The shift in $a$ absorbs the term in the fluctuation of the NSNS two-form $\delta P[B_2]=\rmd P[\iota_\Phi B_2]+P[\iota_\Phi H_3]$. The bosonic part of the vector and hypermultiplet actions equals
\begin{equation}
 \begin{aligned}
 S_B^{(2)}=\int\operatorname{vol}_g\,\biggl[{}
 & \frac12 F_{ab}F^{ab} - 4D_a\bar\varphi D^a\varphi + 2M\bar\varphi\varphi\\
 & + 16F_{ab}\left(\bar\varphi T^{ab} + \varphi\bar T^{ab}\right) + 64\bar\varphi^{\,2}T_{ab}T^{ab} + 64\varphi^{\,2}\bar T_{ab}\bar T^{ab}\\
 & + \sum_{\alpha=1}^2\left(\partial_a\bar q_\alpha\partial^a q_\alpha
       +\frac{\mathcal R+M}{4}\bar q_\alpha q_\alpha\right)\biggr]\,.
 \end{aligned}
 \label{d3:boson-action}
\end{equation}
Here $\mathcal R$ is the scalar curvature of the induced metric. The scalar $M$, the two-forms $T,\bar T$ and the connection in the covariant derivative $D$ are fixed background fields. In terms of $D=\rmd-\rmi r_{\rm wv} A_R$, the vector-multiplet scalars have charges $r_{\rm wv}(\varphi)=-2$ and $r_{\rm wv}(\bar\varphi)=2$. Using the twisted one-forms $\mathcal D\chi=\rmd\chi+\rmi(\nu_F-\frac14)\rmd\hat\tau$ and $\mathcal D\phi=\rmd\phi+\frac{\rmi}{2}\rmd\hat\tau$, their covariant derivatives are
\begin{equation}
 \begin{aligned}
 D\varphi&=(\rmd+2\rmi A_R)\varphi\,,\qquad
 D\bar\varphi=(\rmd-2\rmi A_R)\bar\varphi\,,\\
 A_R&=-\frac{\cos^2\theta\,\mathcal D\chi+\sin^2\theta\,\mathcal D\phi}{6W}
       -\frac{\rmi}{W}\rmd\hat\tau
       +\frac{\rmi}{2}\left(\frac14+\nu_F\right)\rmd\hat\tau\,.
 \end{aligned}
 \label{d3:normal-connection}
\end{equation}
This $\mathrm U(1)_R$ is the R symmetry of the worldvolume $\mathcal N=2$ multiplets, not the boundary generator $R$. The background scalar and curvature evaluated on the light cycle are
\begin{equation}
 M=-\frac{12+49\sin^2\theta+11\sin^4\theta}{4L^2W^{5/2}}\,,\qquad
 \mathcal R=\frac{52+27\sin^2\theta+9\sin^4\theta}{4L^2W^{5/2}}\,.
 \label{d3:geometric-masses}
\end{equation}

The tensor backgrounds are obtained from pulling back the three-forms 
\begin{equation}
 \begin{aligned}
 T&=-\frac1{16}\left(P[\iota_{n_5-\rmi n_6}(H_3+\rmi F_3)]\right)^+\,,\qquad \star T=T\,,\\
 \bar T&=\frac1{16}\left(P[\iota_{n_5+\rmi n_6}(H_3-\rmi F_3)]\right)^-\,,\qquad \star\bar T=-\bar T\,.
 \end{aligned}
 \label{d3:linear-twoforms}
\end{equation}
To write these fields explicitly on the worldvolume, we first introduce the worldvolume frames
\begin{equation}
 \begin{aligned}
 \vartheta^0&=LW^{1/4}\rmd\hat\tau\,,&
 \vartheta^1&=\sqrt{\frac23}\,LW^{1/4}\rmd\theta\,,\\
 \vartheta^2&=\frac{4L}{3W^{3/4}}
       (\cos^2\theta\,\mathcal D\chi+\sin^2\theta\,\mathcal D\phi)\,,&
 \vartheta^3&=\sqrt{\frac23}\,\frac{L\sin\theta\cos\theta}{W^{3/4}}
       (2\mathcal D\chi-\mathcal D\phi)\,.
 \end{aligned}\label{eq:wvframes}
\end{equation}
Since the tensors are (anti-)self-dual, we also introduce the (anti-)self-dual forms
\begin{equation}
	 \Sigma_i^\pm=\vartheta^0\wedge\vartheta^i
       \pm\frac12\epsilon_{ijk}\,\vartheta^j\wedge\vartheta^k\,,\qquad
 \epsilon_{123}=1\,,\qquad \star\Sigma_i^\pm=\pm\Sigma_i^\pm
\end{equation}
in terms of which
\begin{equation}
 \begin{aligned}
 T&=-\frac{e^{\rmi(\phi+\chi)}}{16LW^{1/4}}
       \left[\sqrt3\,(\rmi\Sigma_1^++\Sigma_3^+)
       +\frac{\sqrt2\sin\theta\cos\theta}{W}\,\Sigma_2^+\right]\,,\\
 \bar T&=-\frac{e^{-\rmi(\phi+\chi)}}{16LW^{1/4}}
       \left[\sqrt3\,(-\rmi\Sigma_1^-+\Sigma_3^-)
       +\frac{\sqrt2\sin\theta\cos\theta}{W}\,\Sigma_2^-\right]\,.
 \end{aligned}
 \label{d3:explicit-twoforms}
\end{equation}
Their contractions take much simpler forms
\begin{equation}
	 T_{ab}T^{ab}=\frac{\sin^2\theta\cos^2\theta}{32L^2W^{5/2}}e^{2\rmi(\phi+\chi)}\,,\qquad
 \bar T_{ab}\bar T^{ab}=\frac{\sin^2\theta\cos^2\theta}{32L^2W^{5/2}}e^{-2\rmi(\phi+\chi)}\,.
\end{equation}

\subsubsection{Fermionic fluctuations}

As we show below, the fermionic fluctuations have the quadratic action \eqref{d3:fermion-background-action}. We derive it from the Green--Schwarz action for the D3-brane fermions. In the doubled-spinor convention of section~\ref{d3:kappa}, we write $\Theta=(\vartheta,\vartheta^c)^T$, with Pauli matrices $s_i$ acting on the doublet. At quadratic order, the Green–-Schwarz action is
\begin{equation}
 S_F^{(2)}=\frac{\pi}{(2\pi\ell_s)^4g_s}\int\operatorname{vol}_g\,
 \Theta^T\mathcal B_\Theta(1+\Gamma_\kappa)\mathscr D_F\Theta\,,\qquad
 \mathcal B_\Theta=s_1\otimes C\,,
 \label{d3:fermion-action}
\end{equation}
where $C$ is the ten-dimensional charge conjugation matrix. The ten-dimensional kinetic operator is 
\begin{equation}
 \begin{aligned}
 \mathscr D_F={}&\Gamma^a\nabla_a^{(10)} - \frac18\Gamma^a H_{aMN}\Gamma^{MN}s_1 - \frac18 \Gamma^a\left(\slashed F_3s_2 + \frac{\rmi}{2}\slashed{\widetilde F}_5s_3\right)\Gamma_a + \frac14\left(\slashed H_3s_1-\slashed F_3s_2\right)\,.
 \end{aligned}
 \label{d3:fermion-operator}
\end{equation}
Here $\slashed H_3=H_{ABC}\Gamma^{ABC}/3!$, and similarly for the other forms. The index $a$ is tangent to the brane and $M,N$ are ambient indices. The connection $\nabla^{(10)}$ includes the normal connection, and the five-form is the gauge-invariant $\widetilde F_5$ of~\eqref{sugra:polarization}.

We impose the IIB chirality condition and the physical $\kappa$- symmetry gauge
\begin{equation}
 \Gamma_*\Theta=\Theta\,,\qquad
 \Gamma_*=-\rmi\Gamma_1\cdots\Gamma_{10}\,,\qquad
 \Gamma_\kappa\Theta=-\Theta\,.
 \label{d3:euclidean-fermion-chirality}
\end{equation}
Pulled back onto the D3 light giant, the fermionic kinetic operator reduces to
\begin{equation}
 \mathscr D_F=\Gamma^a\nabla_a^{(10)}+\mathcal M_F\,,\qquad
 \mathcal M_F=-\frac14\left(\slashed H_3s_1+\slashed F_3s_2\right)
 -\frac{\rmi}{16}\Gamma^a\slashed{\widetilde F}_5\Gamma_as_3\,.
 \label{d3:fermion-flux-matrix}
\end{equation}

To decompose $\mathscr D_F$, we use the worldvolume frames \eqref{eq:wvframes} and the normal frames $(n_1,\ldots,n_6)$, ordered as in~\eqref{d3:proper-fluctuations}. We split the ten-dimensional $\Gamma$-matrices as
\begin{equation}
 \begin{gathered}
\gamma^a=\begin{pmatrix}0&\sigma^a\\\bar\sigma^a&0\end{pmatrix}\,,\qquad
 \sigma^0=\bar\sigma^0=\mathbf1_2\,,\qquad
 \sigma^i=-\rmi\tau_i\,,\quad\bar\sigma^i=\rmi\tau_i\,,\\
 \gamma_5=-\gamma^0\gamma^1\gamma^2\gamma^3
          =\operatorname{diag}(\mathbf1_2,-\mathbf1_2)\,,\qquad
 \Gamma_{t_a}=\gamma^a\otimes\mathbf1_8\,,\quad
 \Gamma_{n_j}=\gamma_5\otimes\rho_j\,.
 \end{gathered}
 \label{d3:fermion-clifford-split}
\end{equation}
The $\tau_i$ are Pauli matrices. We take the six normal Clifford matrices to be
\begin{equation}
 \begin{aligned}
 \rho_1&=\tau_1\otimes\mathbf1_2\otimes\mathbf1_2\,,&
 \rho_2&=\tau_2\otimes\mathbf1_2\otimes\mathbf1_2\,,\\
 \rho_3&=\tau_3\otimes\tau_1\otimes\mathbf1_2\,,&
 \rho_4&=\tau_3\otimes\tau_2\otimes\mathbf1_2\,,\\
 \rho_5&=\tau_3\otimes\tau_3\otimes\tau_1\,,&
 \rho_6&=\tau_3\otimes\tau_3\otimes\tau_2\,.
 \end{aligned}
 \label{d3:normal-clifford-basis}
\end{equation}

The full spinor $\Theta$ can be split according to its IIB doublet, the four-dimensional tangent spinor, and the eight-dimensional normal spinor. Splitting the tangent spinor further into its two Weyl components gives four blocks of independent components, ordered by $(s_3,\gamma_5)=(+,+),(+,-),(-,+),(-,-)$. In this basis the chirality matrices are
\begin{equation}
	\rho_*=-\rmi\rho_1\cdots\rho_6=-\tau_3\otimes\tau_3\otimes\tau_3\,,\quad \text{and} \quad \Gamma_*=\mathbf1_2\otimes\gamma_5\otimes\rho_*\,,
\end{equation}
and $\Gamma_\kappa=-s_3\otimes\gamma_5\otimes\mathbf1_8$. The physical $\kappa$-symmetry gauge remove half of the components, so we decompose the full spinor as
\begin{equation}
 \begin{gathered}
 \Theta=\mathcal N_\Theta
 \begin{pmatrix}\Xi_+\\0_{16}\\0_{16}\\\Xi_-\end{pmatrix}\,,\qquad
 \mathcal N_\Theta=\frac{e^{-\rmi\pi/4}}{2\sqrt{T_3}}\,,\qquad
 T_3=\frac{1}{(2\pi)^3g_s\ell_s^4}\,,
 \end{gathered}
 \label{d3:full-spinor-split}
\end{equation}
and the $\Xi_{\pm}$ spinor components are
\begin{equation}
 \begin{gathered}
 \Xi_+=2\lambda_1\otimes|+-+\rangle
       +2\lambda_2\otimes|-++\rangle
       +\psi_1\otimes|++-\rangle
       +\psi_2\otimes|---\rangle\,,\\
 \Xi_-=2\bar\lambda_1\otimes|+--\rangle
       +2\bar\lambda_2\otimes|-+-\rangle
       +\bar\psi_1\otimes|+++\rangle
       +\bar\psi_2\otimes|--+\rangle\,,
 \end{gathered}
 \label{d3:explicit-fermion-decomposition}
\end{equation}
where $\lambda_A,\bar\lambda_A$ are the gaugini, with $A$ the $\mathrm{SU}(2)_R$ index, and $\psi_I,\bar\psi_I$ are the hyperini, with $I$ the symplectic hypermultiplet index. The basis vectors are $|+\rangle=(1,0)^T$ and $|-\rangle=(0,1)^T$, and the three $\pm$ labels give the eigenvalues of the three normal $\tau_3$ factors. To distinguish the spinor sectors, we take $\rho_1\rho_2\rho_3\rho_4=-\tau_3\otimes\tau_3\otimes\mathbf1_2$ and the internal rotation generator $r_{\rm wv}=\rmi\rho_5\rho_6=-\mathbf1_2\otimes\mathbf1_2\otimes\tau_3$, for which the eigenvalues are
\begin{equation}
 \begin{array}{c|ccccc}
 &s_3&\gamma_5&\rho_*&\rho_1\rho_2\rho_3\rho_4&r_{\rm wv}\\ \hline
 \lambda_A     &+1&+1&+1&+1&-1\\
 \bar\lambda_A &-1&-1&-1&+1&+1\\
 \psi_I        &+1&+1&+1&-1&+1\\
 \bar\psi_I    &-1&-1&-1&-1&-1
 \end{array}
 \label{d3:fermion-chirality-assignments}
\end{equation}
Using this split of the fermions, we can decompose the couplings so that the fermions complete the bosonic sector above into full supersymmetric multiplets. The ten-dimensional covariant derivative splits as
\begin{equation}
 \nabla_a^{(10)}=\nabla_a^{(4)}
       +\frac14\omega^\perp_{aij}\Gamma^{n_i n_j}\,.
 \label{d3:fermion-connection-split}
\end{equation}
The normal part of the spin-connection combines with the five-form contribution into a non-trivial connection on the spinors. Their covariant derivatives are 
\begin{equation}
	D_a \Psi =( \nabla_a^{(4)} + q_R \rmi A_{R,a}) \Psi\,,\quad \text{with} \quad  q_R(\lambda_A) = -q_R(\bar\lambda_A) = -q_R(\psi_I) = q_R(\bar\psi_I) = 1\,,
\end{equation}
where $q_R=-r_{\rm wv}$ relative to the convention $D=\rmd-\rmi r_{\rm wv}A_R$ above, and $A_R$ is given in~\eqref{d3:normal-connection}. The three-form flux contributions evaluate to
\begin{equation}
 P[\iota_{n_5+\rmi n_6}(H_3+\rmi F_3)]=0\,,\qquad
 P[\iota_{n_5-\rmi n_6}(H_3-\rmi F_3)]=0\,,
 \label{d3:fermion-flux-polarizations}
\end{equation}
and
\begin{equation}
 \begin{aligned}
 -\frac18P[\iota_{n_5-\rmi n_6}(H_3+\rmi F_3)]_{ab}\sigma^{ab}
     &=2T_{ab}\sigma^{ab}\,,\\
 \frac18P[\iota_{n_5+\rmi n_6}(H_3-\rmi F_3)]_{ab}\bar\sigma^{ab}
     &=2\bar T_{ab}\bar\sigma^{ab}\,,
 \end{aligned}
 \label{d3:fermion-tensor-reduction}
\end{equation}
where we define the standard Weyl matrices $\sigma^{ab}=\tfrac12(\sigma^a\bar\sigma^b-\sigma^b\bar\sigma^a)$ and $\bar\sigma^{ab}=\tfrac12(\bar\sigma^a\sigma^b-\bar\sigma^b\sigma^a)$. 

Putting the terms together, we find that the quadratic fermionic action reduces to
\begin{equation}
 \begin{aligned}
 S_F^{(2)}=\int\operatorname{vol}_g\,\biggl[{}
 &-2\rmi\lambda^A\sigma^aD_a\bar\lambda_A - \frac{\rmi}{2}\bar\psi^I\bar\sigma^aD_a\psi_I + \frac{\rmi}{2}T_{ab}\psi^I\sigma^{ab}\psi_I - \frac{\rmi}{2}\bar T_{ab}\bar\psi^I\bar\sigma^{ab}\bar\psi_I
 \biggr]\,,
 \end{aligned}
 \label{d3:fermion-background-action}
\end{equation}
which precisely matches the fermionic actions of an abelian vector multiplet and a hypermultiplet~\cite{Hama:2012bg}, supplemented by an additional $\mathrm U(1)_R$ connection.

\subsubsection{Supersymmetric quadratic action}
\label{d3:supersymmetric-quadratic}

The bosonic and fermionic fluctuations therefore assemble into a rigid supersymmetric theory on the induced geometry. We use the already mentioned index conventions: $a,b=0,1,2,3$ are tangent-frame indices, $\alpha=1,2$ labels the complex scalars $q_\alpha$, $A,B=1,2$ are $\mathrm{SU}(2)_R$ indices, and $I,J=1,2$ are symplectic hypermultiplet indices. We raise doublet indices as $\lambda^A=\epsilon^{AB}\lambda_B$ and $\psi^I=\Omega^{IJ}\psi_J$, with $\epsilon^{12}=\Omega^{12}=1$ and $\epsilon_{12}=\Omega_{12}=-1$. The quadratic D3-brane action can then be written as
\begin{equation}
 \begin{aligned}
 S_{\rm D3}^{(2)}&=S_B^{(2)}+S_F^{(2)}=\int\operatorname{vol}_g\,
                 (\mathcal L_{\rm v}+\mathcal L_{\rm h})\,,\\[2pt]
 \mathcal L_{\rm v}={}&\frac12F_{ab}F^{ab}
       -4D_a\bar\varphi D^a\varphi+2M\bar\varphi\varphi
       -2\rmi\lambda^A\sigma^aD_a\bar\lambda_A\\
 &+16F_{ab}(\bar\varphi T^{ab}+\varphi\bar T^{ab})
       +64\bar\varphi^{\,2}T_{ab}T^{ab}
       +64\varphi^{\,2}\bar T_{ab}\bar T^{ab}\,,\\[2pt]
 \mathcal L_{\rm h}={}&\sum_{\alpha=1}^2\left(\partial_a\bar q_\alpha\partial^aq_\alpha
       +\frac{\mathcal R+M}{4}\bar q_\alpha q_\alpha\right)
       -\frac{\rmi}{2}\bar\psi^I\bar\sigma^aD_a\psi_I\\
 &+\frac{\rmi}{2}T_{ab}\psi^I\sigma^{ab}\psi_I
       -\frac{\rmi}{2}\bar T_{ab}\bar\psi^I\bar\sigma^{ab}\bar\psi_I\,.
 \end{aligned}
 \label{d3:full-supersymmetric-action}
\end{equation}
The background fields are those in~\eqref{d3:normal-connection}, \eqref{d3:geometric-masses} and~\eqref{d3:explicit-twoforms}. Since the action is now written manifestly in terms of a vector multiplet and a hypermultiplet, the global supersymmetry transformations in \cite{Hama:2012bg} also hold on the D3-brane worldvolume, at least to the quadratic order studied here.

\subsubsection{One-loop determinant}

We compute the one-loop determinant of the worldvolume action \eqref{d3:full-supersymmetric-action} from the single-letter indices of its multiplets, and obtain \eqref{d3:determinant-split} at generic fugacities. The action contains an abelian $\mathcal N=2$ vector multiplet and a full hypermultiplet neutral under the gauge group. Their standard single-letter indices are~\cite{Gadde:2011uv}
\begin{equation}
 \begin{aligned}
 f_{\rm v}(p,q,t)&=-\frac p{1-p}-\frac q{1-q}
                  +\frac{pq/t-t}{(1-p)(1-q)}\,,\\
 f_{\rm h}(p,q,t;a_{\rm h})&=
       \frac{(\sqrt t-pq/\sqrt t)(a_{\rm h}+a_{\rm h}^{-1})}{(1-p)(1-q)}\,.
 \end{aligned}
 \label{d3:standard-multiplet-letters}
\end{equation}
Here $p,q$ are associated with the worldvolume rotations, $t$ with the worldvolume R symmetry, and $a_{\rm h}$ is the hypermultiplet flavour fugacity. We denote the worldvolume rotation Cartans by $j_{1,\rm wv},j_{2,\rm wv}$, the $\mathrm{SU}(2)_{R,\rm wv}$ Cartan by $R_{\rm wv}$, the $\mathrm U(1)_{r,\rm wv}$ charge by $r_{\rm wv}$, and twice the $\mathrm{SU}(2)_{F,\rm wv}$ Cartan by $F_{\rm wv}$. The letter of the individual fields is given by
\begin{equation}
 w_{\rm wv}=p^{j_{1,\rm wv}+j_{2,\rm wv}+r_{\rm wv}/2}
 q^{j_{2,\rm wv}-j_{1,\rm wv}+r_{\rm wv}/2}
 t^{R_{\rm wv}-r_{\rm wv}/2}a_{\rm h}^{F_{\rm wv}}\,.
 \label{d3:standard-worldvolume-grading}
\end{equation}
The two supersymmetric derivatives have $(j_{1,\rm wv},j_{2,\rm wv})=(\tfrac12,\tfrac12)$ and $(-\tfrac12,\tfrac12)$, giving weights $p$ and $q$, respectively. The bottom scalar letters have vanishing worldvolume spin, and their charges are 
\begin{equation}
 \begin{array}{c|ccc|c}
 \text{Letter}&R_{\rm wv}&r_{\rm wv}&F_{\rm wv}&\text{Weight}\\ \hline
 \bar\varphi&0&2&0&pq/t\\
 q_1&\tfrac12&0&1&\sqrt t\,a_{\rm h}\\
 \bar q_2&\tfrac12&0&-1&\sqrt t/a_{\rm h}
 \end{array}
 \label{d3:worldvolume-field-charges}
\end{equation}
To relate the worldvolume weights to the boundary weights, we write the two angular Killing vectors of the worldvolume as
\begin{equation}
 K_{1,\rm wv}=-\frac12(\partial_\phi+\partial_\chi)\,,\qquad
 K_{2,\rm wv}=\frac12(\partial_\chi-\partial_\phi)\,.
 \label{d3:worldvolume-lifted-generators}
\end{equation}
The corresponding charges are $j_{1,\rm wv},j_{2,\rm wv}$. The boundary Killing vectors~\eqref{sugra:generators}, which defines the boundary partition function, decompose into these worldvolume Killing vectors as
\begin{equation}
 \begin{aligned}
 K_R&=\frac12\partial_\chi-\partial_\phi
 =\frac12K_{1,\rm wv}+\frac32K_{2,\rm wv}\,,\quad  K_F=-\partial_\chi
 =K_{1,\rm wv}-K_{2,\rm wv}\,,
 \end{aligned}
 \label{d3:worldvolume-charge-generators}
\end{equation}
and the boundary charges decompose into worldvolume charges as
\begin{equation}
 \begin{aligned}
 R=\frac12j_{1,\rm wv}+\frac32j_{2,\rm wv}-\frac14r_{\rm wv}\,,\quad 
 F=j_{1,\rm wv}-j_{2,\rm wv}-\frac12r_{\rm wv}\,,\quad j_2&=R_{\rm wv}\,,\quad j_1&=\frac12F_{\rm wv}\,.
 \end{aligned}
 \label{d3:worldvolume-boundary-charge-map}
\end{equation}
The protected states contributing to the partition function saturate the BPS bound~\eqref{qft:supercharge} and are graded by
\begin{equation}
 \begin{aligned}
 w&=x^{2D-R}u^F=x^{2(2j_2+R)}u^F = x^{j_{1,\rm wv}+3j_{2,\rm wv}+4R_{\rm wv}-r_{\rm wv}/2}
    u^{j_{1,\rm wv}-j_{2,\rm wv}-r_{\rm wv}/2}\,.
 \end{aligned}
 \label{d3:protected-letter-weight}
\end{equation}
Comparing with~\eqref{d3:standard-worldvolume-grading} fixes the worldvolume fugacities in boundary variables:
\begin{equation}
 p_{\rm wv}=x^2\,,\qquad q_{\rm wv}=\frac xu\,,\qquad
 t_{\rm wv}=x^4\,,\qquad a_{\rm h}=1\,.
 \label{d3:worldvolume-index-fugacities}
\end{equation}
In terms of the boundary fugacities, the single-letter indices of the vector multiplet and the hypermultiplet on the brane reduce to
\begin{equation}
 \begin{aligned}
 i_{\rm v}(x,u)&=\frac{(x^{-2}+x^2)x/u}{1-x/u}
                  -\frac{x^2+x^4}{1-x^2}\,,\\
 i_{\rm h}(x,u)&=\frac{2x^2}{1-x^2}-\frac{2x/u}{1-x/u}\,.
 \end{aligned}
 \label{d3:multiplet-letters}
\end{equation}
The single-letter index of the brane is their sum,
\begin{equation}
 i_{\rm D3}=i_{\rm v}+i_{\rm h}
 =\frac{x^2+(x^{-1}-2x)/u}{1-x/u}\,.
 \label{d3:character}
\end{equation}
At generic fugacities, the nonzero-mode Gaussian is the plethystic exponential of $i_{\rm D3}$ times the Casimir factor,
\begin{equation}
 \begin{aligned}
 Z_{\rm nz}&=e^{-\beta E_c^{\rm D3}}P(x,u)\,,\\
 P(x,u)&\equiv\PE[i_{\rm D3}(x,u)]
 =\frac{(x/u;x/u)_\infty^2}{(x^2;x/u)_\infty(1/(xu);x/u)_\infty}\,.
 \end{aligned}
 \label{d3:determinant-split}
\end{equation}
Here the Casimir energy on the brane, computed with a $\zeta$-function regulator, equals
\begin{equation}
 E_{c}^{\rm D3} = \frac12\left(1-\frac2{1+2\nu_F}\right)\,.
 \label{d3:zeta-vacuum}
\end{equation}

\subsection{Comparison with the boundary partition function}\label{d3:ensembles}

We first compare the results of the brane quantization directly with the grand canonical ensemble of the field theory. With the brane Casimir factor included, the results \eqref{d3:grand-contribution} and \eqref{d3:grand-unflavoured} below directly match the grand canonical corrections \eqref{qft:leadingrefined} at $y = 1$ and \eqref{qft:leadingunrefined}, respectively. Both matches hold before the sum over large gauge images. The inverse Laplace transform then recovers the corresponding canonical answer. 

We use the boundary condition of section \ref{sugra:action} that fixes the electric potential. On the slice $y=1$, with $x=e^{-\beta/2}$ and $u=e^{\beta\nu_F}$,
\begin{equation}
 \mu=\frac{2\beta L^4}{27\pi g_s\ell_s^4}(1-4\nu_F^2)\,,\qquad
 \mathcal C=\frac{\beta(1-4\nu_F^2)}8\,.
 \label{d3:ensemble-dictionary}
\end{equation}
The classical action~\eqref{d3:classical-weight} and the Casimir energy~\eqref{d3:zeta-vacuum}, with $\nu_F\mapsto\sigma\nu_F$ for the two light giants, combine to give
\begin{equation}
 \begin{aligned}
 S_{\rm D3}^\sigma(\mu)+\beta E_c^{\sigma,\rm D3}
 &=\frac{2\mu}{1+2\sigma\nu_F}
   +\frac\beta2\left(1-\frac2{1+2\sigma\nu_F}\right)\,.
 \end{aligned}
 \label{d3:ensemble-actions}
\end{equation}
We denote by $Z_{\rm D3}^{\rm np}$ the leading non-perturbative D3 correction relative to the perturbative bulk saddle at fixed bulk potential. Before summing the large gauge images, the saddle partition function is $Z_{\rm IIB}^{\rm p}[1+Z_{\rm D3}^{\rm np}+\cdots]$. We now use the field-theory potentials $\tau=\rmi\beta/(2\pi)$ and $\varphi=-2\tau\nu_F$, with $x=e^{\rmi\pi\tau}$ and $u=e^{\rmi\pi\varphi}$. Combining~\eqref{d3:ensemble-actions} with the one-loop determinant~\eqref{d3:determinant-split} gives precisely the $y=1$ specialization of~\eqref{qft:leadingrefined}: 
\begin{equation}
 \begin{aligned}
 Z_{\rm D3}^{\rm np}(\mu;x,u)
 &=\sum_{\sigma=\pm1}
   \rme^{-\frac{2\tau\mu}{\tau-\sigma\varphi}}
   (xu^\sigma)^{-\tau/(\tau-\sigma\varphi)}
   \frac{(x/u^\sigma;x/u^\sigma)_\infty^2}
        {(x^2;x/u^\sigma)_\infty(1/(xu^\sigma);x/u^\sigma)_\infty}\,.
 \end{aligned}
 \label{d3:grand-contribution}
\end{equation}
Summing the two flavour sectors before taking $u\to1$ gives, in agreement with~\eqref{qft:leadingunrefined},
\begin{equation}
	Z_{\rm D3}^{\rm np}(\mu;x,1) = -\frac{2\rmi\mu}{\pi\tau}e^{-2\mu}\,.
 \label{d3:grand-unflavoured}
\end{equation}
The canonical corrections follow from the inverse Laplace transform of the full perturbative saddle times the D3 contribution:
\begin{equation}
 \delta Z_N^{\rm D3}(x,u)
 =\int_{c-\rmi\infty}^{c+\rmi\infty}\frac{\rmd \mu}{2\pi\rmi}\,
 \rme^{-N\mu+J^{\rm p}(\mu;x,1,u)}Z_{\rm D3}^{\rm np}(\mu;x,u)\,.
 \label{d3:inverse-laplace}
\end{equation}
Here $J^{\rm p}$ is given in~\eqref{qft:refinedmodified}, and $c\in\mathbb R$.\footnote{In this expression the sum over large gauge transformations has been absorbed into the choice of contour.} The two corresponding canonical non-perturbative corrections are
\begin{equation}
 \begin{aligned}
 \frac{\delta Z_N^{\rm D3}(x,u)}{Z_N^{\rm p}(x,1,u)}
 &=\sum_{\sigma=\pm1}(xu^\sigma)^N
   \frac{(x/u^\sigma;x/u^\sigma)_\infty^2}
        {(x^2;x/u^\sigma)_\infty(1/(xu^\sigma);x/u^\sigma)_\infty}\,,\\[3pt]
 \frac{\delta Z_N^{\rm D3}(x,1)}{Z_N^{\rm p}(x,1,1)}
 &=-(N+2)x^{N+1}\,.
 \end{aligned}
 \label{d3:canonical-contribution}
\end{equation}
With $Z_N^{\rm p}(x,1,u)=e^{-\mathcal C N^2}\mathcal I_\infty(x,1,u)$ and $P(x,u^\sigma)=\mathcal G_\sigma(x,1,u)$, the flavoured result reproduces~\eqref{qft:leadingfixedrank} at $y=1$. Its unflavoured limit agrees with the leading fixed-rank correction~\cite{Bourdier:2015wda}. Interestingly, we find that the inverse Laplace transform also directly reproduces the term $+2$ in the prefactor $-(N+2)$.

It would be interesting to derive the prefactor $N+2$ directly in the canonical ensemble from the D3-brane worldvolume theory. At $u=1$, the light-giant embeddings $a z_1+b z_2=0$ are parametrized by $[a:b]\in\mathbf{CP}^1$. Quantizing this collective coordinate, including the effect of the worldvolume fluctuations, may account for the finite shift.

For the half-BPS ABJM index, it was suggested in \cite{Gautason:2025plx} that dual giant gravitons play a role in the grand canonical ensemble. We leave for future work the question of whether dual D3-brane giants play an analogous role for the LS index and the $\mathcal N=4$ Schur index.

\section{Conclusion}
\label{sec:conclusion}

The Leigh--Strassler fixed point provides a setting with a reduced amount of supersymmetry in which to test finite-$N$ holography and the relation between canonical and grand canonical ensembles. Its Pilch-Warner dual has a deformed internal geometry and nontrivial three-form flux. On the boundary side, the supersymmetric $S^3\times S^1$ partition function remains accessible through a Fermi-gas description. 

We find a full bulk match for the perturbative part of the partition function in both ensembles. Depending on the coupling of the topological contributions to the supergravity on-shell action, and one-loop fluctuations, we reproduce the boundary result at fixed rank $N$ or at fixed chemical potential $\mu$. 

We also find a direct match for the leading non-perturbative D3-brane contribution in the grand canonical ensemble. At angular fugacity $y=1$, the classical giant-graviton actions and their one-loop determinants reproduce the leading non-perturbative correction to the boundary partition function in this ensemble, and the inverse Laplace transform recovers the corresponding finite-$N$ correction. In the context of the Mellin-Barnes representation in figure \ref{fig:mellincontour} we have thus explicitly reproduced the zero-pole, associated to the supergravity on-shell action, and the pole at $-1$ associated to the leading giant graviton. Reproducing the higher order poles we will leave for future work. It is interesting to see that the non-perturbative sector of D3-brane giants in fact knows about the supergravity on-shell action coming from the zero-pole through such a Mellin-Barnes representation.

In the free-gas limit, the LS partition function can be exactly mapped onto the $\mathcal N=4$ Schur index, upon a redefinition of parameters, even though the bulk geometries do not resemble each other. It would be interesting to investigate whether these geometries are related by a supersymmetric $Q$-exact deformation, which could explain the equality of the corresponding partition functions. 

A complementary question concerns the microscopic origin of the ensemble prescription. It would be particularly interesting to derive the appropriate boundary terms directly from a worldsheet calculation in type IIB string theory. Such a derivation would give a microscopic foundation for the ensemble choice used here. It would also clarify how the canonical and grand canonical descriptions arise from the underlying string path integral.

\noindent \textbf{Acknowledgments} 

\noindent We thank Pieter-Jan de Smet for discussions and early collaboration in this project. JvM is supported by the STFC Consolidated Grant ST/X000575/1 and the UK Engineering and Physical Sciences Research council grant number EP/Z000106/1. JvM is grateful for the continuous hospitality of the ITF at the KU Leuven. 

\appendix
\section{Special functions and the diagonal interaction factor}
\label{app:specialfunctions}

We collect the special-function conventions used in section~\ref{sec:qft} and give the diagonal part of the interaction factor~\eqref{qft:interactionweight}. We use Jacobi theta functions with $x=e^{\rmi\pi\tau}$, $|x|<1$, defined by
\begin{equation}
 \begin{aligned}
 \vartheta_1(z,x)&=2\sum_{n=0}^{\infty}(-1)^n x^{(n+1/2)^2}\sin[(2n+1)z]\,,\\
 \vartheta_2(z,x)&=2\sum_{n=0}^{\infty} x^{(n+1/2)^2}\cos[(2n+1)z]\,,\\
 \vartheta_3(z,x)&=1+2\sum_{n=1}^{\infty}x^{n^2}\cos(2nz)\,,\\
 \vartheta_4(z,x)&=1+2\sum_{n=1}^{\infty}(-1)^n x^{n^2}\cos(2nz)\,.
 \end{aligned}
 \label{app:thetaseries}
\end{equation}
The fractional powers are taken on the branch fixed by $x=e^{\rmi\pi\tau}$. For $0<x<1$ they are positive. The Pochhammer symbol and elliptic gamma function are
\begin{equation}
 (a;t)_\infty=\prod_{n=0}^{\infty}(1-at^n)\,, \qquad \Gamma_e(z;s,t)=\prod_{m,n=0}^{\infty} \frac{1-s^{m+1}t^{n+1}/z}{1-zs^mt^n}\,,
 \label{app:ellipticgamma}
\end{equation}
with $|s|,|t|<1$. When $|st|<|z|<1$ the latter has the logarithmic expansion
\begin{equation}
 \log\Gamma_e(z;s,t) =\sum_{k=1}^{\infty} \frac{z^k-(st/z)^k}{k(1-s^k)(1-t^k)}\,.
 \label{app:gammalog}
\end{equation}
In terms of the single-letter functions $f_{\rm LS}$ and $f_0$ of~\eqref{qft:letter} and~\eqref{qft:interaction}, the diagonal oscillator factors of the LS theory and its free Fermi-gas contribution are
\begin{equation}
 \begin{split}
 \exp\!\left[\sum_{k=1}^{\infty}\frac{f_{\rm LS}(x^k,y^k,u^k)}{k}\right] ={}&(x^2y;x^2y)_\infty(x^2/y;x^2/y)_\infty\\
 &\times\Gamma_e(xu;x^2y,x^2/y) \Gamma_e(x/u;x^2y,x^2/y)\,,\\[3pt]
 \exp\!\left[\sum_{k=1}^{\infty}\frac{f_0(x^k,u^k)}{k}\right] ={}&\frac{(x^2;x^2)_\infty^3}{\vartheta_4(\pi\varphi/2,x)}\,, \qquad u=e^{\rmi\pi\varphi}\,.
 \end{split}
 \label{app:diagonalfactors}
\end{equation}
The first expression is the one-loop product of the vector multiplet and the two chiral multiplets in the localization formula~\cite{Assel:2014paa}. Their ratio gives $e^A$, where $A=\sum_{k\geq1}\delta f(x^k,y^k,u^k)/k$ is the diagonal interaction in~\eqref{qft:interactionweight}:
\begin{equation}
 \begin{split}
 e^A={}&\frac{(x^2y;x^2y)_\infty(x^2/y;x^2/y)_\infty \vartheta_4(\pi\varphi/2,x)}{(x^2;x^2)_\infty^3}\\
 &\times\Gamma_e(xu;x^2y,x^2/y) \Gamma_e(x/u;x^2y,x^2/y)\,.
 \end{split}
 \label{app:diagonalproduct}
\end{equation}
There are $N$ diagonal terms in the matrix integral, so their total contribution is $e^{NA}$, as in~\eqref{qft:fermiprefactor}. At $y=1$ the two single-letter functions coincide, thus $e^A=1$. 

\section{The Pilch--Warner background in explicit coordinates}
\label{app:backgroundcoordinates}

This appendix gives the coordinate expressions for the background in section~\ref{sugra:background}, using the angular parametrization~\eqref{sugra:embedding-coordinates} introduced in section~\ref{sugra:supersymmetry}. In these coordinates $W=1+\sin^2\theta$, and the internal left-invariant one-forms are
\begin{equation}
 \sigma_1-\rmi\sigma_2 =e^{-\rmi\xi_3}\left(\rmd\xi_1+\rmi\sin\xi_1\,\rmd\xi_2\right)\,, \qquad \sigma_3=\rmd\xi_3+\cos\xi_1\,\rmd\xi_2\,.
 \label{sugra:forms}
\end{equation}
In these coordinates, the one-forms of \eqref{sugra:embedding-oneforms} and the internal three-form $\Omega_3$, are
\begin{equation}
 k_z=\frac12\cos^2\theta\,\sigma_3\,,\qquad k_w=-\sin^2\theta\,\rmd\phi\,,\qquad \Omega_3=-\sigma_1\wedge\sigma_2\wedge\sigma_3\,.
 \label{sugra:angular-oneforms}
\end{equation}
The ten-dimensional metric is $\rmd s_{10}^2=L^2\sqrt W(\rmd s_{\mathrm{AdS}_5}^2+\rmd\tilde s_5^2)$. For the external factor we use the Euclidean global coordinates of section~\ref{sugra:background},
\begin{equation}
 \begin{aligned}
 \rmd s_{\mathrm{AdS}_5}^2&=\rmd\rho^2+\cosh^2\rho\,\rmd\hat\tau^2 +\sinh^2\rho\,\rmd\Omega_3^2\,,\\
 \rmd\Omega_3^2&=\rmd\eta^2+\sin^2\eta\,\rmd\chi_1^2 +\cos^2\eta\,\rmd\chi_2^2\,,\\
 \operatorname{vol}_{\mathrm{AdS}_5}&=\cosh\rho\,\sinh^3\rho\, \rmd\rho\wedge\rmd\hat\tau\wedge\operatorname{vol}_{S^3}\,.
 \end{aligned}
 \label{sugra:global-ads-coordinates}
\end{equation}
Here $\operatorname{vol}_{S^3}$ is the volume form of the unit external three-sphere. The internal metric is
\begin{equation}
 \begin{aligned}
 \rmd\tilde s_5^2={}&\frac23\rmd\theta^2 + \frac{\cos^2\theta}{3W}\left[\sigma_1^2+\sigma_2^2 + \frac{4W}{5W-2}\sigma_3^2\right] + \frac{\cos^2\theta\sin^2(2\theta)}{18W^2(5W-2)} \left(\sigma_3+\frac{5W-2}{\cos^2\theta}\rmd\phi\right)^2\,.
 \end{aligned}
 \label{sugra:angular-metric}
\end{equation}
The axion and dilaton are $C_0=0$ and $e^\Phi=g_s$. With the RR normalization of section~\ref{sugra:background}, the form potentials are
\begin{equation}
 \begin{aligned}
 B_2-\rmi C_2&=\frac{L^2}{3}\sqrt{\frac23}\,e^{-\rmi\phi} \left[\cos\theta\,\rmd\theta+\rmi\frac{\cos^2\theta\sin\theta}{W}(\sigma_3-\rmd\phi)\right] \wedge(\sigma_1-\rmi\sigma_2)\,,\\
 C_4^{\mathrm{el}}&=-\rmi L^4\sinh^4\rho\, \rmd\hat\tau\wedge\operatorname{vol}_{S^3}\,,\\
 C_4^{\mathrm{mag}}&=-\frac{4L^4\cos^4\theta}{27W}\,\rmd\phi\wedge\Omega_3\,.
 \end{aligned}
 \label{sugra:angular-fluxes}
\end{equation}
The volume form of the unit round five-sphere becomes
\begin{equation}
 \omega_5=\frac18\cos^3\theta\sin\theta\, \rmd\theta\wedge\rmd\phi\wedge\Omega_3\,, \qquad \int_{S^5}\omega_5=\pi^3\,.
 \label{sugra:angular-volume}
\end{equation}
The gauge-invariant five-form \eqref{sugra:polarization}, with $H_3=\rmd B_2$ and $F_3=\rmd C_2$ has components
\begin{equation}
 \begin{aligned}
 \widetilde F_5^{\mathrm{el}}=F_5^{\mathrm{el}} &=-4\rmi L^4\cosh\rho\,\sinh^3\rho\, \rmd\rho\wedge\rmd\hat\tau\wedge\operatorname{vol}_{S^3} =-4\rmi L^4\operatorname{vol}_{\mathrm{AdS}_5}\,,\\
 \widetilde F_5^{\mathrm{mag}}&=\frac{16L^4\cos^3\theta\sin\theta}{27W^2} \rmd\theta\wedge\rmd\phi\wedge\Omega_3\,.
 \end{aligned}
 \label{sugra:angular-fiveform}
\end{equation}

\subsection{Killing spinors and the supersymmetric circle}
\label{app:explicitspinor}

We write the Killing spinors of the background in \cite{Pilch:2000ej} in our conventions, as an ordered product of exponentials acting on a constant spinor. To specify their components, choose the external orthonormal frames
\begin{equation}
 \begin{gathered}
 e^1=L W^{1/4}\cosh\rho\,\rmd\hat\tau\,,\qquad e^5=L W^{1/4}\rmd\rho\,,\\
 (e^2,e^3,e^4)=L W^{1/4}\sinh\rho\, (\rmd\eta,\sin\eta\,\rmd\chi_1,\cos\eta\,\rmd\chi_2)\,,
 \end{gathered}
 \label{app:spinor-external-frame}
\end{equation}
and the internal frames
\begin{equation}
 \begin{aligned}
 e^6&=L\sqrt{\frac23}W^{1/4}\rmd\theta\,,& e^7-\rmi e^8&=\frac{L\cos\theta}{\sqrt3 W^{1/4}} e^{-\rmi\phi}(\sigma_1-\rmi\sigma_2)\,,\\
 e^9&=L\sqrt{\frac23}\frac{\sin\theta\cos\theta}{W^{3/4}} (\sigma_3-\rmd\phi)\,,& e^{10}&=\frac{2L}{3W^{3/4}} (\cos^2\theta\,\sigma_3+2\sin^2\theta\,\rmd\phi)\,.
 \end{aligned}
 \label{app:spinor-internal-frame}
\end{equation}
The $\Gamma$-matrices are given in Euclidean signature, with $\Gamma_{ab}=\Gamma_a\Gamma_b$ for $a\ne b$ and $\Gamma_*=-\rmi\Gamma_1\cdots\Gamma_{10}$. The IIB doublet of spinors is $\epsilon=(\varepsilon,\varepsilon^c)^T$, with Pauli matrices $s_i$ acting on its two entries. The angle $\alpha(\theta)$ in~\eqref{d3:spinor-factors} is defined continuously on $0\leq\theta\leq\pi/2$ by
\begin{equation}
 \cos\alpha=\frac{\cos\theta}{\sqrt W}\,,\qquad \sin\alpha=\frac{\sqrt2\sin\theta}{\sqrt W}\,,\qquad \alpha(0)=0\,,\quad\alpha(\pi/2)=\frac\pi2\,.
 \label{app:spinor-angle}
\end{equation}
Up to a constant normalization, the full ten-dimensional Killing spinor is
\begin{equation}
 \begin{aligned}
 \epsilon={}& \exp\!\left[-\frac{\rmi\pi}{4}s_3\right] W^{1/8}\cdot \exp\!\left[\frac{\alpha(\theta)}2 \bigl(\sqrt3\,\Gamma_{9\,10}+s_1\Gamma_{7\,10}\bigr)\right]\\
 &\cdot\exp\!\left[\frac{\rmi\rho}{2}s_3\Gamma_{5\,10}\right]\cdot \exp\!\left[-\frac\eta2\Gamma_{25}\right]\cdot \exp\!\left[-\frac{\chi_1}2\Gamma_{32}\right]\cdot \exp\!\left[-\frac{\chi_2}2\Gamma_{45}\right]\\
 &\cdot\exp\!\left[\frac{\rmi\hat\tau}2s_3\Gamma_{1\,10}\right]\cdot \exp\!\left[\rmi\phi s_3\right]\epsilon_0\,,
 \end{aligned}
 \label{app:coordinate-killing-spinor}
\end{equation}
where $\epsilon_0$ is independent of all ten coordinates and obeys
\begin{equation}
 \Gamma_*\epsilon_0=\epsilon_0\,,\qquad \Gamma_{69}\epsilon_0=-\rmi s_3\epsilon_0\,,\qquad \Gamma_{78}\epsilon_0=-\rmi s_3\epsilon_0\,.
 \label{app:background-spinor-projections}
\end{equation}
The projections leave eight complex parameters in the independently doubled description, corresponding to eight real supercharges. 

The supercharge chosen in~\eqref{qft:supercharge} is selected by
\begin{equation}
 s_3\epsilon_0=-\epsilon_0\,,\qquad \rmi s_3\Gamma_{1\,10}\epsilon_0=-\epsilon_0\,,\qquad \Gamma_{32}\epsilon_0=\rmi\epsilon_0\,,\qquad \Gamma_{45}\epsilon_0=\rmi\epsilon_0\,.
 \label{app:index-spinor-projections}
\end{equation}
Together with~\eqref{app:background-spinor-projections}, these conditions leave one complex parameter. They give $\partial_{\hat\tau}\epsilon=-\epsilon/2$, $\partial_{\chi_a}\epsilon=-\rmi\epsilon/2$ and $\mathcal L_{K_R}^{}\epsilon=\rmi\epsilon$, while $\mathcal L_{K_F}^{}\epsilon=0$. Thus $\mathcal L_Y^{}\epsilon=0$ precisely when the twists satisfy~\eqref{sugra:twist-supersymmetry}. In the twisted coordinates~\eqref{sugra:periodic-coordinates} the spinor is independent of $\hat\tau$. 

\section{Five-dimensional derivation of the supersymmetric Casimir energy}
\label{app:casimir}

In this appendix we compute the finite boundary contribution to the classical on-shell action used in section~\ref{sugra:action}. The five-dimensional Chern--Simons coupling fixes the coefficient of the flavour-dependent term, which we evaluate on the thermal background with a supersymmetric boundary prescription. The unflavoured contribution follows from~\cite{BenettiGenolini:2016tsn}. 

We first compute the fixed-flux action $S_{\mathrm{ren}}^{(+)}=S_N^{\mathrm{cl}}$. In the last subsection we take its Legendre transform to $S_{\mathrm{ren}}^{(-)} = S_\mu^{\mathrm{cl}}$. The signs $(\pm)$ refer to $\alpha=\pm1$ in~\eqref{sugra:ensemble-action}.

\subsection{The five-dimensional theory and the mixed Chern--Simons coefficient}
\label{app:casimir:truncation}

To keep a non-trivial flavour source, we need a vector multiplet in addition to the graviphoton of minimal gauged supergravity. We therefore use a $\mathbb Z_2\times\mathbb Z_2$ invariant $\mathcal N=2$ sector of the $S^5$ reduction to $SO(6)$ gauged supergravity, with three vectors $A^I$~\cite{Khavaev:2000gb,Bobev:2014jva}. For the Pilch-Warner background, with vector-multiplet scalars $\alpha_{\rm v},\eta_{\rm v}$ and hypermultiplet scalars $(\chi_{\rm h},\vartheta_{\rm h})$, the Euclidean bosonic action is
\begin{equation}
 \begin{aligned}
 S_5={}&\frac{1}{4\pi G_5}\int_{M_5}\dd^5x\sqrt{g_5}\,
 \biggl[-\frac14R_5+\frac14\sum_I e_I F^I_{MN}F^{I\,MN}
 +3(\partial\alpha_{\rm v})^2+(\partial\eta_{\rm v})^2
 +\frac12(\partial\chi_{\rm h})^2\\
 &\hspace{34mm}
 +\frac18\sinh^2(2\chi_{\rm h})(D\vartheta_{\rm h})^2
 +\mathcal P\biggr]
 +\frac{\rmi}{2\pi G_5}\int_{M_5}A^3\wedge F^1\wedge F^2\,.
 \end{aligned}
 \label{app:casimir:action}
\end{equation}
Here $I=1,2,3$, $F^I=\dd A^I$, $(e_1,e_2,e_3)=(e^{4\alpha_{\rm v}-4\eta_{\rm v}}, e^{4\alpha_{\rm v}+4\eta_{\rm v}},e^{-8\alpha_{\rm v}})$, $D\vartheta_{\rm h}=\dd\vartheta_{\rm h}+g(A^1+A^2-A^3)$, and $\mathcal P$ is the scalar potential, with $\mathcal P_*=-3/L^2$ at the LS vacuum. Thus the gravitational terms there are $-(R_5+12/L^2)/(16\pi G_5)$. The normalization is $\kappa_5^2=8\pi G_5=54\pi^4g_s^2\ell_s^8/L^5$. At the LS point, $e^{6\alpha_{{\rm v}*}}=2$, $e^{2\chi_{{\rm h}*}}=3$, $\eta_{{\rm v}*}=0$, $L=3/(2^{2/3}g)$, where $g$ is the five-dimensional gauge coupling. The radius $L$ is the same as in section~\ref{sugra:background}, and flux quantization gives $1/G_5=N^2g^3/(4\pi)$, with $N$ the Page charge~\eqref{sugra:page-rank}. For the charges $R(X,Y,Z)=(1/2,1/2,1)$, $F(X,Y,Z)=(1,-1,0)$ and the Higgsed direction $\mathcal B=g(A^1+A^2-A^3)$, the basis
\begin{equation}
 gA^1=\frac12\mathcal A_R+\mathcal A_F+\frac16\mathcal B\,,\qquad gA^2=\frac12\mathcal A_R-\mathcal A_F+\frac16\mathcal B\,,\qquad gA^3=\mathcal A_R-\frac23\mathcal B
 \label{app:casimir:gauge-basis}
\end{equation}
diagonalizes the vector kinetic terms at the LS point: $\mathcal A_R$ and $\mathcal A_F$ are massless, and $\mathcal B$ gets a mass from $\vartheta_{\rm h}$ with $m_{\mathcal B}^2L^2=6$. We write $\mathcal F_R=\dd\mathcal A_R$ and $\mathcal F_F=\dd\mathcal A_F$. 

The coefficient needed for the finite flavour term follows from the Chern--Simons coupling. Setting $\mathcal B=0$ and using $1/(2\pi G_5g^3)=N^2/(8\pi^2)$ gives
\begin{equation}
 S_{\mathrm{CS}} =\frac{\rmi N^2}{8\pi^2} \int_{M_5}\mathcal A_R\wedge \left(\frac14\mathcal F_R\wedge\mathcal F_R -\mathcal F_F\wedge\mathcal F_F\right)\,,
 \label{app:casimir:bulk-CS}
\end{equation}
This gives $k_{RRR}=\frac34N^2$, $k_{RFF}=-N^2$ and $k_{RRF}=k_{FFF}=0$ in the convention $S_{\mathrm{CS}}=\rmi\int\mathcal I_5$, $\dd\mathcal I_5=k_{IJK}\mathcal F_I\wedge\mathcal F_J\wedge\mathcal F_K/(24\pi^2)$. 

We now specify the thermal background in these variables. In the gauge corresponding to the twisted identification of section~\ref{sugra:supersymmetry}, the R and flavour connections and their source parameters are
\begin{equation}
 \begin{aligned}
     \mathcal A_R =& -\rmi\nu_R\dd\hat\tau\,,\qquad \mathcal A_F=-\rmi\nu_F\dd\hat\tau\,,
 \end{aligned}
 \label{app:casimir:euclidean-conventions}
\end{equation}
and the twist parameters are
\begin{equation}
    u = e^{\rmi\oint\mathcal A_F}=e^{\beta\nu_F}=e^{\rmi\pi\varphi}\,,\qquad
     b_a=1+\epsilon_a\,,\qquad \nu_R=\frac{3-b_1-b_2}{2}\,.
\end{equation}
Thus $\nu_F=\rmi\pi\varphi/\beta$ on the chosen branch of the flavour holonomy, and the term $-16\nu_F^2$ below is the $16\pi^2\varphi^2/\beta^2$ term in~\eqref{qft:hopfcasimir}.

We also set $\eta_{\rm v}=0$ and $\mathcal B=0$ on this filling. Switching on $\nu_F$ is locally a flavour gauge transformation of the unflavoured solution, so $\mathcal F_F=0$. The flavour connection nevertheless has the holonomy $u$ in~\eqref{app:casimir:euclidean-conventions}. The bulk density, Gibbons--Hawking term, covariant counterterms and Chern--Simons term are then independent of $\nu_F$, so $S_{\rm cov}(\nu_F)=S_{\rm cov}(0)$ in any covariant scheme, and the flavour dependence comes entirely from the finite subtraction restoring supersymmetry.

\subsection{The finite mixed counterterm}
\label{app:casimir:mixed-counterterm}

The bulk coefficient $k_{RFF}$ fixes the normalization, but computing the finite term also requires a supersymmetric boundary prescription. We use the component construction of~\cite{Closset:2019ucb}, restricted to our bosonic thermal background. To state the result, first rescale the boundary metric so that $\partial_{\hat\tau}$ has unit norm, with $\hat\tau\sim\hat\tau+\beta$, and write
\begin{equation}
 \dd s_{4,\mathrm{unit}}^2 = (\dd\hat\tau+c)^2 + \dd s_3^2\,.
 \label{app:casimir:circle-metric}
\end{equation}
Here $c$ is the one-form describing the angular twist, and $g_3$ is the metric on the three-dimensional space $M_3$ obtained by reducing along the thermal circle. Both are determined by the boundary metric. The flavour-dependent finite term to add to the action is
\begin{equation}
 \Delta S_F=\frac{\beta k_{RFF}\nu_F^2}{32\pi^2}
 \int_{M_3}\dd^3x\sqrt{g_3}\,
 \left[R_3+8\nu_R^2+(*_3\dd c)^2-4\nu_R\,c\cdot(*_3\dd c)\right]\,.
 \label{app:casimir:geometric-integral}
\end{equation}
Here $R_3$ and $*_3$ are the Ricci scalar and Hodge star of $g_3$, and all contractions use this metric. The fields required by supersymmetry have already been substituted, leaving only the metric and the chemical potentials in the integrand.\footnote{With $\beta=2\pi\beta_{S^1}$ in the notation of~\cite{Closset:2019ucb}, the mixed term obtained from their equations (3.106) and (3.118) reduces to $\mathcal W_{RFF}=\beta k_{RFF}\sigma_F^2\int\sqrt{g_3}[(R_3-2H^2+2V^2)/4-a_R\cdot V]/(8\pi^2)$. Substituting $H=2\rmi\nu_R$, $V=-\rmi *_3\dd c$, $a_R=Hc/2+3V/4$ and $\sigma_F=-\rmi\nu_F$ gives~\eqref{app:casimir:geometric-integral}, with $\Delta S_F=-\mathcal W_{RFF}$ to cancel the anomalous supersymmetry variation.} Equation~\eqref{app:casimir:geometric-integral} is the value of the prescribed boundary term on the flat-flavour thermal family. Thus the calculation consists of reading $k_{RFF}$ from~\eqref{app:casimir:bulk-CS} and evaluating the geometric integral in~\eqref{app:casimir:geometric-integral}.

\subsection{Evaluation on the twisted thermal saddle}
\label{app:casimir:evaluation}

We now determine $g_3$ and $c$ explicitly. Introduce boundary coordinates $0\leq\vartheta\leq\pi/2$ and $\varphi_i\sim\varphi_i+2\pi$ by
\begin{equation}
 \vartheta=\frac\pi2-\eta\,,\qquad
 \varphi_i=-\widetilde\chi_i\,,\qquad k_i=-\rmi\epsilon_i\,,
 \qquad b_i=1+\rmi k_i=1+\epsilon_i\,.
 \label{app:casimir:angular-dictionary}
\end{equation}
The twisted boundary metric of section~\ref{sugra:supersymmetry} gives
\begin{equation}
    \begin{aligned}
        \dd s_4^2 =& \,\dd\hat\tau^2 + \dd\vartheta^2 + \cos^2\vartheta (\dd\varphi_1 + k_1\dd\hat\tau)^2  + \sin^2\vartheta (\dd\varphi_2 + k_2\dd\hat\tau)^2\,,\\
        \dd s_{4,\mathrm{unit}}^2= &\, p^{-1}\dd s_4^2\,, \qquad \dd s_3^2 =p^{-1}\left(\dd\vartheta^2+\cos^2\vartheta\,\dd\varphi_1^2 +\sin^2\vartheta\,\dd\varphi_2^2\right) -c^2\,,
    \end{aligned}
\end{equation}
where
\begin{equation}
p = 1 + k_1^2 \cos^2\vartheta + k_2^2 \sin^2\vartheta \,, \qquad c=\frac{k_1\cos^2\vartheta\,\dd\varphi_1 +k_2\sin^2\vartheta\,\dd\varphi_2}{p}\,.
 \label{app:casimir:unit-circle}
\end{equation}
We first evaluate the integral for real $k_i$ and then continue analytically to the desired twists. To simplify the expressions we introduce $t=\sin^2\vartheta$, $A=1+k_1^2$ and $B=1+k_2^2$, so that $ p = A+(B-A)t$. The quantities entering~\eqref{app:casimir:geometric-integral} are
\begin{gather}
\int\dd\varphi_1\dd\varphi_2\,\sqrt{g_3}\,\dd\vartheta=\frac{2\pi^2\dd t}{p^2}\,,\qquad
 R_3=\frac{20AB}{p}-6(A+B)-2\,,\nonumber\\
 (*_3\dd c)^2=-4+\frac{4AB}{p}\,,\qquad
 c\cdot(*_3\dd c)=-2k_1k_2\,.
 \label{app:casimir:geometric-data}
\end{gather}
Using $\int_0^1\dd t/p^2=1/(AB)$, $\int_0^1\dd t/p^3=(A+B)/(2A^2B^2)$ and $\nu_R=(3-b_1-b_2)/2$, the integral becomes
\begin{equation}
 \begin{aligned}
 &\int_{M_3}\dd^3x\sqrt{g_3}\, \left[R_3+8\nu_R^2+(*_3\dd c)^2-4\nu_R\,c\cdot(*_3\dd c)\right]\\
 &\qquad=2\pi^2\int_0^1\dd t\, \left[\frac{24AB}{p^3} + \frac{8\nu_R^2+8\nu_R k_1k_2-6(A+B)-6}{p^2}\right] = 8\pi^2\frac{b_1+b_2}{b_1b_2}\,.
 \end{aligned}
 \label{app:casimir:integral-value}
\end{equation}
Substituting $k_{RFF}=-N^2$ in~\eqref{app:casimir:geometric-integral} therefore gives
\begin{equation}
 \Delta S_F=-\beta\,\frac{N^2(b_1+b_2)}{4b_1b_2}\nu_F^2\,.
 \label{app:casimir:flavour-action}
\end{equation}
For the round background, $c=0$, $R_3=6$ and $\nu_R=1/2$, so the integrand in~\eqref{app:casimir:geometric-integral} is simply $8$. With $\operatorname{Vol}(S^3)=2\pi^2$, this gives $\Delta S_F=-\beta N^2\nu_F^2/2$ directly.

The unflavoured supersymmetric action from~\cite{BenettiGenolini:2016tsn} is $S_N^{\rm cl}|_{\nu_F=0}=\beta N^2(b_1+b_2)^3/(64b_1b_2)$. Subtracting the minimally renormalized action $81\beta N^2/512$ and adding~\eqref{app:casimir:flavour-action} gives the full finite boundary contribution on our background,
\begin{equation}
 \Delta S_{\mathrm{susy}}=\beta N^2
 \left[\frac{(b_1+b_2)\bigl[(b_1+b_2)^2-16\nu_F^2\bigr]}{64b_1b_2}
 -\frac{81}{512}\right]\,.
 \label{app:casimir:finite-completion}
\end{equation}
This is the term added to minimal subtraction in the fixed-flux ensemble, in agreement with~\eqref{sugra:finite-completion}. The resulting on-shell action is
\begin{equation}
 S_N^{\rm cl}=\beta E_{\rm susy}^{\rm cl}\,,\qquad E_{\rm susy}^{\rm cl} =\frac{N^2(b_1+b_2)}{64b_1b_2} \left[(b_1+b_2)^2-16\nu_F^2\right]\,,
 \label{app:casimir:energy}
\end{equation}
which reproduces~\eqref{sugra:flavoured-coefficient} and \eqref{sugra:canonical-action}, with $S_N^{\rm cl}=S_{\mathrm{ren}}^{(+)}=\mathcal C N^2$. 

\subsection{Supersymmetric volume and the electric potential}
\label{app:casimir:electric-potential}

The same finite boundary term also renormalizes the electric source
$\mu$ conjugate to $N$. The grand canonical on-shell action is obtained
by the Legendre transform $S_{\mathrm{ren}}^{(-)}=S_{\mathrm{ren}}^{(+)}-N\mu$. The conjugate variable is the holonomy of the four-form gauge potential~\cite{vanMuiden:2026lno}, and it can be computed from
\begin{equation}
 \mu=\left[\mu_{\mathrm{bare}} + \partial_N S_{\partial}^{(+)} \right]_{\mathrm{finite}}\,,\qquad \mu_{\mathrm{bare}}=\frac{\rmi}{(2\pi)^3g_s\ell_s^4}\int C_4^{\mathrm{el}}\,,
 \label{app:casimir:canonical-source-shift}
\end{equation}
where $S_\partial^{(+)}=S_{\mathrm{GHY}}+S_{\mathrm{ct,min}}+\Delta S_{\mathrm{susy}}$ is the boundary functional, with the finite supersymmetric counterterm given in the previous subsection. In terms of the bulk length scales and the string coupling, this becomes
\begin{gather}
 \mu =\frac{L^4}{2\pi^3g_s\ell_s^4} \left[\int_{\mathrm{AdS}_5}\rmd^5x\sqrt{g_{\mathrm{AdS}_5}} +\frac{27\pi^4g_s^2\ell_s^8}{2L^8}S_{\partial}^{(+)} \right]_{\mathrm{finite}} =\frac{L^4}{2\pi^3g_s\ell_s^4} \operatorname{Vol}_{\mathrm{AdS}_5}^{\mathrm{SUSY}}\,,
 \label{app:casimir:volume-period-consistency}
\end{gather}
which agrees with the volume definition in~\eqref{sugra:renormalized-volume}. At fixed boundary sources, $\mathcal C$ is independent of $N$, so
$\mu=\partial_N S_N^{\mathrm{cl}}=2\mathcal C N$.
The Legendre transform therefore gives
\begin{equation}
 S_\mu^{\mathrm{cl}}=S_N^{\mathrm{cl}}-N\mu = -\frac{\mu^2}{4\mathcal C}\,,
 \label{app:casimir:euclidean-ensembles}
\end{equation}
as in~\eqref{sugra:flavoured-casimir}.

\section{The heavy giant at $w=0$}
\label{app:heavygiant}

We now consider the heavy embedding $w=0$ introduced in section~\ref{d3:embedding}. Its classical weight is $x^{2N}$, equal to that of a pair of light giants. We show that it preserves the same supersymmetry as the light giants, but its one-loop determinant has unpaired fermionic zero modes. Because of these zero modes, the isolated heavy-giant Gaussian gives no contribution at classical weight $x^{2N}$ when $y = 1$. We use the Euclidean conventions of sections~\ref{d3:kappa} and~\ref{d3:one-loop}, with $b_1=b_2=1$, $x=e^{-\beta/2}$, $y=1$ and $u=e^{\beta\nu_F}$.

\subsection{Embedding, geometry and classical action}

The brane is stabilised on the cycle parametrised by $(\hat\tau,\xi_1,\xi_2,\xi_3)$, sitting at $w=0$ $(\theta=0)$ and $\rho=0$. Here the one-forms $e^6$ and $e^9$ vanish on the worldvolume, and the nonvanishing pullbacks of the frames~\eqref{app:spinor-external-frame}--\eqref{app:spinor-internal-frame} are
\begin{equation}
 P[e^1]=L\,\rmd\hat\tau\,,\qquad
 P[e^7]-\rmi P[e^8]=\frac{L}{\sqrt3}\,e^{-\rmi\phi}(\sigma_1-\rmi\sigma_2)\,,\qquad
 P[e^{10}]=\frac{2L}{3}\,\sigma_3\,.
 \label{hg:frame}
\end{equation}
The induced metric on the brane is a squashed sphere times Euclidean time,
\begin{equation}
 \rmd s^2_{\rm D3}=L^2\Bigl[\rmd\hat\tau^2+\frac13\bigl(\sigma_1^2+\sigma_2^2\bigr)+\frac49\,\sigma_3^2\Bigr]\,,\qquad
 \sqrt{\det h}=\frac29L^4\sin\xi_1\,,
 \label{hg:metric}
\end{equation}
where $\Omega_3=-\sigma_1\wedge\sigma_2\wedge\sigma_3=\sin\xi_1\,\rmd\xi_1\wedge\rmd\xi_2\wedge\rmd\xi_3$ has integral $16\pi^2$ in the coordinate orientation. 

The two-form potentials don't contribute to the worldvolume since
\begin{equation}
 B_2-\rmi C_2=\frac1{\sqrt3}\,(e^6+\rmi e^9)\wedge(e^7-\rmi e^8)\,.
 \label{hg:twoform-frame}
\end{equation}

Integrating the volume gives $S_{\rm DBI}=\beta\Delta$ with
\begin{equation}
 \Delta=\frac{4L^4}{9\pi g_s\ell_s^4}=\frac{3N}{2}\,,\qquad (R,F)=(N,0)\,,\qquad
 S_{\rm D3}^{w}=\beta\Bigl(\Delta-\frac R2\Bigr)=\beta N\,,\qquad e^{-S^w_{\rm D3}}=x^{2N}\,.
 \label{hg:action}
\end{equation}
The brane charges $(R,F)$ follow from the prescription of section~\ref{d3:action}. As for the light cycles, $\Delta=3R/2$, which is a result of supersymmetry.

\subsection{Kappa symmetry}

This subsection shows that the heavy giant preserves the same supersymmetry as the light giants. The worldvolume set of frame labels is now $\{1,7,8,10\}$. The $\kappa$-projector now equals
\begin{equation}
 \Gamma_{(4)}=-\Gamma_1\Gamma_7\Gamma_8\Gamma_{10}=-\Gamma_{78}\Gamma_{1\,10}\,,\qquad
 \Gamma_\kappa=s_3\Gamma_{(4)}\,,\qquad\Gamma_\kappa^2=1\,.
 \label{hg:projector}
\end{equation}

Compared with the light cycle, two entries of the commutation relations used in~\eqref{d3:projector-commutation} change. As before, $\Gamma_{25}$, $\Gamma_{32}$, $\Gamma_{45}$ and $\Gamma_{1\,10}$ still commute with $\Gamma_\kappa$ and $s_3\Gamma_{5\,10}$ still anticommutes with it. Now, however, $\Gamma_{9\,10}$ and $s_1\Gamma_{7\,10}$ both anticommute with $\Gamma_\kappa$, so $\mathcal N$ no longer commutes with the projector. On the cycle this is not important, since $\alpha(0)=0$ and $\mathcal U_1=e^{-\rmi\pi s_3/4}$ at $\theta=0$. The reduction~\eqref{d3:kappa-reduced} and the localisation argument of section~\ref{d3:kappa} then apply without change: supersymmetry requires $\rho=0$, where the condition is the constant projection $\Gamma_\kappa\epsilon_0=\epsilon_0$. Using $\Gamma_{78}\epsilon_0=-\rmi s_3\epsilon_0$ from~\eqref{app:background-spinor-projections},
\begin{equation}
 \Gamma_\kappa\epsilon_0=-s_3\Gamma_{1\,10}\Gamma_{78}\epsilon_0=\rmi\Gamma_{1\,10}\epsilon_0\,,\qquad
 (\rmi\Gamma_{1\,10}-1)\epsilon_0=0\,.
 \label{hg:constant-spinor}
\end{equation}

As for the light giants in section~\ref{d3:kappa}, the projection leaves four complex parameters in the Euclidean doubled-spinor description. The heavy giant is therefore half-BPS. 

\subsection{Quadratic fluctuations}

For the four scalars along AdS we denote with \(X^I\) and the internal complex scalar $\zeta=\sqrt{2/3}\,L\eta$ $(\eta=\delta w)$, the bosonic quadratic action on the brane is 
\begin{equation}
\begin{aligned}
\frac{S_B^{(2)}}{T_3}
={}&\frac12\int\operatorname{vol}_h\!
\left[\sum_{I=1}^4\left((\partial X^I)^2+\frac{(X^I)^2}{L^2}\right)
+\left|\rmd\zeta-\tfrac{\rmi}{3}\sigma_3\zeta\right|^2-\frac7{L^2}|\zeta|^2+\tfrac12\mathcal F_{1\,ab}\mathcal F_1^{ab}\right]\\
&-\rmi\int C_1\wedge f+\frac{2L^4}{9}
\int(\bar\eta\,\rmd\eta-\eta\,\rmd\bar\eta)\wedge\Omega_3 .
\end{aligned}
\label{hg:boson-action}
\end{equation}
The last term in this quadratic action arises from the four-form gauge potential in a regular gauge. The vector fluctuations are $a=2\pi\ell_s^2\delta A_{\rm wv}$, and \(\mathcal F_1=\rmd a - B_1\), where the two-form variations are
\begin{equation}
    B_1-\rmi C_1=\tfrac{\sqrt6}{9}L^2(\rmd\eta+\rmi\sigma_3\eta)\wedge(\sigma_1-\rmi\sigma_2)\,.
\end{equation}
and subsequently cause a mixing between the scalar $\eta$ and the two vector components along the $S^2$ base of the three-sphere.

The fermionic action is \(S_F^{(2)}=\tfrac{T_3}{2}\int\operatorname{vol}_h\, \Theta^T\mathcal B_\Theta(1+\Gamma_\kappa)\mathscr D_F^{\,w}\Theta\), with \(\Gamma_*\Theta=\Theta\) and \(\Gamma_\kappa\Theta=-\Theta\) as in \eqref{d3:fermion-action} and \eqref{d3:euclidean-fermion-chirality}. For convenience we define the vectors dual to the frames \eqref{sugra:forms} as
\begin{equation}
    E_1=\cos\tilde{\xi}_3\,\partial_{\xi_1}+\sin\tilde{\xi}_3\,V,\qquad E_2=\sin\tilde{\xi}_3\,\partial_{\xi_1}-\cos\tilde{\xi}_3\,V,\qquad E_3=\partial_{\tilde{\xi}_3}\,, 
\end{equation}
with $V=(\partial_{\tilde{\xi}_2}-\cos\xi_1\,\partial_{\tilde{\xi}_3})/\sin\xi_1$. In the frame \(\tilde\phi=0\), the operator \eqref{d3:fermion-flux-matrix} including the flavour twist is
\begin{equation}
\begin{aligned}
L\mathscr D_F^{\,w}={}&
\Gamma_1\!\left(\partial_{\hat\tau}+2\rmi\nu_F\partial_\chi+\tfrac{\rmi}{2}E_3
+\tfrac{\rmi}{4}(\Gamma_{69}+\Gamma_{78})\right)
+\sqrt3(\Gamma_7E_1+\Gamma_8E_2)+\tfrac32\Gamma_{10}E_3\\
&+\tfrac54\Gamma_{78\,10}+\tfrac14\Gamma_{69\,10}+s_3\Gamma_{12345}
+\tfrac{\sqrt3}{4}\!\left[s_1(\Gamma_{68\,10}+\Gamma_{79\,10})
+s_2(\Gamma_{89\,10}-\Gamma_{67\,10})\right].
\end{aligned}
\label{hg:fermion-operator}
\end{equation}

\subsection{Mode spectrum and one-loop determinant}

The vectors $E_i$ act non-trivially on spin-$j>0$ harmonics of the three-sphere, we will only care about $j=0$, and thus can drop the $E_i$ terms. The table below gives the masses of these modes, with multiplicities.
\begin{center}
\begin{tabular}{@{}ll@{}}
\toprule
Fields & \(\omega=D-R/2\) at \(j=0\)\\
\midrule
 AdS scalars \(X^I\) & \(\pm1\) $(\times 4)$\\
  $a_3$ along $S^1\subset S^3$ & \(\pm2\)\\
 Internal scalar \(\eta\) and $a_1,a_2$ along $S^2 \subset S^3$
 & \(\pm1\) $(\times 2)$, \(\pm\sqrt7\) $(\times 2)$\\
Kappa-fixed fermions \(\Theta\) & \(0\) $(\times 4)$, \(\pm1,\ \pm2,\ \pm\sqrt7\) $(\times 2)$\\
\bottomrule
\end{tabular}
\end{center}
Each mode with mass \(\pm\omega\) contributes gives $(2\pi n/\beta)^2+\omega^2$ to the partition function, up to sign. Writing $k=k_n = 2\pi n/\beta$, the total contributions of the modes in the table above is
\begin{equation}
 \frac{\det K_{F,0}(k)}{\det K_{B,0}^{\rm phys}(k)}
 =\frac{k^4(k^2+1)^2(k^2+4)^2(k^2+7)^2}
        {(k^2+1)^6(k^2+4)(k^2+7)^2}
 =\frac{k^4(k^2+4)}{(k^2+1)^4}.
 \label{hg:homogeneous-determinant}
\end{equation}
which vanishes due to the $n=0$ eigenvalues coming from the fermions. 

In contrast, the light-giant fermions in \eqref{d3:character} have weights \((x/u^\sigma)^n\), \(n\geq1\), \(\sigma=\pm1\), and give no such zero at generic \(u\). At \(u=1\) the \(+1\) in the light character \(x^{-1}+1-x\) is a bosonic orientation mode.

\bibliography{references}
\bibliographystyle{JHEP}
\end{document}